\documentclass[usenatbib]{mn2e}
\usepackage{apjfonts}
\usepackage{epsfig}
\usepackage{amssymb}
\usepackage{amsmath}
\usepackage{fixltx2e} 

\newcommand{\etal}{et al.}

\newcommand{\msun}{M_{\sun}}
\newcommand{\lstar}{L_{\ast}}

\newcommand{\fgas}{f_{\rm gas}}

\newcommand{\disksurv}{H09}

\newcommand{\scaleup}{}

\newcommand\plotone[1]
 {\centering \leavevmode \includegraphics[width={0.99\columnwidth}]{#1}}
\newcommand\plotonesmall[1]
 {\centering \leavevmode \includegraphics[width={0.85\columnwidth}]{#1}}
\newcommand\plotonesmallish[1]
 {\centering \leavevmode \includegraphics[width={0.95\columnwidth}]{#1}}
\newcommand{\plotside}[1]
 {\centering \leavevmode \includegraphics[width={0.95\textwidth}]{#1}}
\newcommand{\acknowledgments}{\begin{small}\section*{Acknowledgments}\end{small}}
\newcommand\altaffilmark[1]{$^{#1}$}
\newcommand\altaffiltext[1]{$^{#1}$}
\title[Morphological Transformation in Mergers]{The Effects of Gas on Morphological 
Transformation in Mergers: Implications for Bulge and Disk Demographics}

\author[Hopkins \etal]{
\parbox[t]{\textwidth}{ 
Philip F. Hopkins\altaffilmark{1}\thanks{E-mail:phopkins@astro.berkeley.edu},
Rachel S.\ Somerville\altaffilmark{2,3}, 
Thomas J.\ Cox\altaffilmark{4,5}, 
Lars Hernquist\altaffilmark{4},
Shardha Jogee\altaffilmark{6},  
Dusan Keres\altaffilmark{4},  
Chung-Pei Ma\altaffilmark{1},  
Brant Robertson\altaffilmark{7,8,9}, 
\&\ Kyle Stewart\altaffilmark{10} }
\vspace*{6pt} \\
\altaffiltext{1}{Department of Astronomy, University of California
  Berkeley, Berkeley, CA 94720} \\
\altaffiltext{2}{Space Telescope
  Science Institute, 3700 San Martin Dr., Baltimore, MD 21218}\\
\altaffiltext{3}{Department of Physics and Astronomy, Johns Hopkins
  University, Baltimore, MD 21218}\\
\altaffiltext{4}{Harvard-Smithsonian Center for Astrophysics, 60
  Garden Street, Cambridge, MA 02138} \\
\altaffiltext{5}{W.~M.\ Keck
  Postdoctoral Fellow at the Harvard-Smithsonian Center for
  Astrophysics} \\
\altaffiltext{6}{Department of Astronomy, University
  of Texas at Austin, Austin, TX} \\
\altaffiltext{7}{Kavli Institute for
  Cosmological Physics, the Department of Astronomy and Astrophysics,
  University of Chicago, 933 East 56th Street, Chicago, IL, 60637}\\
\altaffiltext{8}{Enrico Fermi Institute, 5640 South Ellis Avenue,
  Chicago, IL, 60637} \\
\altaffiltext{9}{Spitzer Fellow}\\
\altaffiltext{10}{Center for Cosmology, Department of Physics and
  Astronomy, The University of California at Irvine, Irvine, CA 92697}\\
}

\date{Submitted to MNRAS, January 20th, 2009}
\begin{document}
\maketitle
\label{firstpage}

\begin{abstract}

Transformation of disks into spheroids via mergers is a well-accepted element of
galaxy formation models. However, recent simulations have shown that bulge
formation is suppressed in increasingly gas-rich mergers. We investigate the
global implications of these results in a cosmological framework, using
independent approaches: empirical halo-occupation models (where galaxies are
populated in halos according to observations) and semi-analytic models. In both,
ignoring the effects of gas in mergers leads to the over-production
of spheroids: low and intermediate-mass galaxies are predicted to be
bulge-dominated ($B/T\sim0.5$ at $<10^{10}\,M_{\sun}$, with almost no 
`bulgeless' systems),
even if they have avoided major mergers. 
Including the different physical
behavior of gas in mergers immediately leads to a dramatic change: bulge
formation is suppressed in low-mass galaxies, observed to be gas-rich (giving
$B/T\sim~0.1$ at $<10^{10}\,M_{\sun}$, 
with a number of bulgeless galaxies in good agreement
with observations). Simulations 
and analytic models which neglect the similarity-breaking behavior of
gas have difficulty reproducing the strong observed morphology-mass 
relation. However, the observed
dependence of gas fractions on mass, combined with suppression of bulge
formation in gas-rich mergers, naturally leads to the observed trends.
Discrepancies between observations and models that ignore the role of gas
increase with redshift; in models that treat gas properly, 
galaxies are predicted to be less bulge-dominated at high redshifts, 
in agreement with the observations. 
We discuss implications for the global bulge mass density and 
future observational tests. 

\end{abstract}


\section{Introduction}
\label{sec:intro}

In a $\Lambda$CDM cosmology,
structure grows hierarchically \citep[e.g.][]{whiterees78}, making
mergers and interactions between galaxies an essential and inescapable
process in galaxy formation. Indeed, mergers are widely believed to be
responsible for the morphologies of spheroids \citep{toomre77}, and observations find recent
merger remnants in considerable abundance in the local universe
\citep{schweizer82,LakeDressler86,Doyon94,ShierFischer98,James99,
  Genzel01,tacconi:ulirgs.sb.profiles,
  rj:profiles,dasyra:mass.ratio.conditions} as well as e.g.\ faint
shells and tidal features common around apparently ``normal'' galaxies
\citep{malin80,schweizer80,schweizer96},
which are thought to be signatures of galaxy collisions
\citep[e.g.][]{hernquistquinn88,hernquist.spergel.92}.

This has led to the concern that there may be too {\em many}
mergers to explain the survival and abundance of galactic disks in the
context of our present understanding of galaxy formation
\citep[see e.g.][and references therein]{stewart:mw.minor.accretion,
purcell:minor.merger.thindisk.destruction}. 
Indeed, the difficulty of forming very late type disks with low $B/D$
is a well-known problem in cosmological simulations, 
which tend to consistently overpredict the bulge
fractions of low-mass galaxies. There has been considerable effort to
address the cause of these difficulties, with a great deal of focus on
improving resolution (in order to properly resolve the bulge and disk
components) and introducing feedback prescriptions to prevent
artificial disk over-cooling, clumping, fragmentation, and angular
momentum transport \citep{weil98:cooling.suppression.key.to.disks,
sommerlarsen99:disk.sne.fb,sommerlarsen03:disk.sne.fb,
thackercouchman00,thackercouchman01,abadi03:disk.structure,
governato04:resolution.fx,governato:disk.formation,
robertson:cosmological.disk.formation,
okamoto:feedback.vs.disk.morphology,scannapieco:fb.disk.sims}.
Similar problems are encountered in semi-empirical models, 
that use observational constraints to populate a halo-occupation 
framework, and evolve this forward in time to follow mergers 
\citep[see e.g.][]{stewart:disk.survival.vs.mergerrates}. 
A number of semi-analytic models, which predict galaxy properties 
from simple physical recipes and adopt simplified prescriptions for behavior 
in mergers, allowing for the efficiency of merger-induced 
bulge formation to be adjusted, still find the same difficulty even with 
these degrees of freedom \citep{somerville:sam,somerville:new.sam,
khochfar:sam,granato:sam,croton:sam}. 

More importantly, the problem can be seen empirically without reference 
to these models. Observations find that $\sim5-10\%$ of low or intermediate-mass 
galaxies ($M_{\ast}\lesssim10^{10}\,\msun$ at redshifts $z\sim0.2-1.2$) are in mergers -- 
strongly morphologically disturbed or in 
``major'' similar-mass pairs about to merge (at small scales and small relative velocities)
\citep{bridge:merger.fractions,bridge:merger.fraction.new.prep,kartaltepe:pair.fractions,
conselice:mgr.pairs.updated,conselice:highz.mgr.rate.vs.mass,
jogee:merger.density.08,lotz:merger.fraction,lin:mergers.by.type}. 
This is equal to or higher than the predicted rates from the models above -- 
the problem does not seem to be that $\Lambda$CDM predicts ``too many'' 
mergers. 
The problem is that, even without correcting for the short expected 
duty cycles/lifetimes of this merger phase, these fractions are {\em already} 
higher than the observed fraction of bulge-dominated galaxies at these 
masses \citep{allen:bulge-disk,balcells:bulge.scaling,
dominquez:bulge.disk.sample,weinzirl:b.t.dist}. 
Invoking subsequent accretion to ``re-build'' disks 
will not eliminate massive bulges formed by these events, nor can 
it operate in a shorter time than the duration of the merger itself, and 
correcting for the expected duty cycles/lifetimes of these events, 
a large fraction of these low-mass galaxies should have experienced 
a violent event in the last $\sim10\,$Gyr 
\citep{hopkins:transition.mass,
lotz:merger.selection,conselice:mgr.pairs.updated}. 
Moreover, although there may be some mass-dependence in 
observed merger fractions, it is weak: a factor $\sim2$ 
increase with mass from $\sim10^{9}-10^{11}\,\msun$ 
\citep{zheng:hod.evolution,
conselice:highz.mgr.rate.vs.mass,patton:mgr.rate.vs.rmag,
bundy:merger.fraction.new,bridge:merger.fraction.new.prep,
lopezsanjuan:mgr.rate.pairs,domingue:2mass.merger.mf}. 
Over the same range, there is an order-of-magnitude or more 
change in the fraction of bulge-dominated galaxies. 
As a result, uniformly increasing or decreasing the ``efficiency'' of 
merger-induced bulge formation is insufficient to explain the 
observed trends \citep[see e.g.][]{stewart:disk.survival.vs.mergerrates}. 
In short, the observed mergers at low and intermediate masses 
must somehow be inefficient at making bulges, 
whereas those at high masses must be efficient 
\citep[][]{koda:disk.survival.prescriptions}.
However, dissipationless 
systems with high or low stellar and/or virial masses behave exactly
the same in actual $N$-body simulations of mergers.  They {\em must}
do so --- if gas is not allowed to change the dynamics of the merger,
then the only force is gravity and the system is scale-free (a merger
of two $10^{9}\,\msun$ systems is exactly equivalent to the merger of
two $10^{12}\,\msun$ systems).
Because gravitational
interactions of dissipationless systems are scale-free, it is not possible to design a
physical prescription that does not treat gas differently from stars
(or where the merger efficiency is just a function of mass ratio and
orbital parameters) that can simultaneously match the observed 
merger fractions and $B/T$
distributions at low and high galaxy masses.  
Some other controlling
parameter or physics must work to suppress bulge formation in low-mass
galaxies. 

\citet{springel:spiral.in.merger} and
\citet{robertson:disk.formation} suggested that this parameter may be
galaxy gas fraction:\footnote{In this paper, we use the term ``gas
  fraction'' to refer specifically to the cooled gas fraction in the
  baryonic disk, i.e. $(m_{\rm HI}+m_{\rm H_{2}})/(m_{\rm HI}+m_{\rm
    H_{2}}+m_{\ast})$, where $m_{\rm HI}$, $m_{\rm H_{2}}$, 
    and $m_{\ast}$ are the atomic, molecular, and stellar masses, respectively.} 
    they showed, in idealized merger simulations,
that even major mergers of sufficiently gas-dominated disks could
produce remnants with large disks. This has since been borne out in
cosmological simulations \citep{governato:disk.formation,
  governato:disk.rebuilding}, and may be seen in observations of
high-redshift disks that are perhaps re-forming after mergers
\citep{genzel:highz.rapid.secular,shapiro:highz.kinematics,kassin:tf.evolution,
  vandokkum:tf.evolution,puech:tf.evol,vanstarkenburg:z2.disk.dynamics,
  robertsonbullock:highz.disk.vs.model}. The growing consensus that 
more realistic treatments of star formation and supernovae feedback --
which suppress efficient early star formation -- play a critical
role in enabling disk formation in simulations (see references above)
is connected with this insight into the link between gas
fraction and disk survival.  Because low-mass disks are observed to be
more gas-rich, this may potentially provide the desired breaking of
self-similarity in dissipationless merger histories.

In \citet{hopkins:disk.survival} (hereafter \disksurv), we developed a
physical model to explain this behavior, and used a large suite of
hydrodynamic simulations
\citep[see][]{cox:kinematics,cox:xray.gas,cox:massratio.starbursts,
  robertson:msigma.evolution,robertson:fp} to show that the
suppression of bulge formation in mergers of more gas-rich disks can
be predicted analytically, and is dynamically inevitable owing to
fundamental properties of gravitational physics and collisional
systems.  Stars, being collisionless, violently relax in mergers, as they 
can mix and randomize their orbits by scattering off
the fluctuating potential, even without net torques. Gas, being
collisional, cannot violently relax, but must have its
angular momentum torqued away in order to dissipate and build a bulge
by forming stars in a central starburst. In a merger, this torque is
primarily internal, from stars in the same disk: the passages and
merger of the secondary induce a non-axisymmetric distortion
in the primary stellar and gas disk. The stellar distortion (in a trailing 
resonance and close proximity to the gas response) efficiently 
removes angular momentum from the gas and allows it to fall to 
the center and form a bulge \citep[see also][]{barneshernquist96}. 
Other sources of torque (hydrodynamic shocks or the gravity of the 
secondary galaxy) represent less efficient, higher-order effects 
(and often tend to spin up the disk). 

In the limit, then, of extremely gas-rich disks, there is no
significant stellar feature available to torque
the cold gas in the disk; gas will rapidly re-form a cold disk in the 
background of the relaxing stellar potential. Because this is a purely
gravitational process, the physics and consequences for fixed 
initial conditions are independent of
uncertain physics associated with ``feedback'' from either star
formation and supernovae or accreting black holes (see \disksurv). 
Feedback is important for {\em setting} these initial
conditions -- over cosmological timescales, it can suppress 
star formation efficiencies and lead to retention of large gas reservoirs, 
and move that gas to larger radii; but for a given set of 
gas properties at the time of the merger, \disksurv\ outline a few 
simple equations motivated by well-understood gravitational 
physics that can robustly estimate how much of the gas will preserve 
its angular momentum content in a given merger. 

We endeavor here to study the consequences of these detailed studies 
of disk survival as a function of merger gas content, for the global demographics 
of bulges and disks as a function of galaxy mass and redshift. To do so, 
we utilize both semi-analytic and semi-empirical cosmological models. 
Semi-analytic models predict galaxy properties from first principles 
via physical recipes; while the latest generation of models reproduce 
many observational quantities quite well, there remain significant 
discrepancies and uncertainties. We therefore compare to a 
semi-empirical model, where we assign galaxy properties 
to dark matter (sub)-halos in order to match observational constraints. 
We combine these with the detailed predictions from simulations 
presented in \disksurv, to make predictions for bulge and disk 
populations. We contrast these predictions with the results from implementing 
the ``traditional'' assumptions regarding bulge formation, 
which ignore the unique role of gas as outlined above. 

In \S~\ref{sec:model} we describe our empirical approach for tracking
stellar and gas masses within the framework of our cosmological merger
trees, and briefly summarize our semi-analytic approach. 
In \S~\ref{sec:consequences} we
outline the key predictions of our gas fraction-dependent bulge-formation
recipe and the differences from those of the recipes generally adopted
in the past. We summarize and
discuss other implications in \S~\ref{sec:discuss}.

Throughout, we assume a flat cosmology with matter and dark energy 
densities $\Omega_{\rm M}=0.3$ and $\Omega_{\Lambda}=0.7$, respectively, 
and Hubble constant $H_{0}=70\,{\rm km\,s^{-1}\,Mpc^{-1}}$, but varying these
parameters within the observationally constrained limits \citep{komatsu:wmap5} 
makes little difference.

\section{The Model}
\label{sec:model}

In order to track merger histories with as few assumptions as possible, 
we construct the following semi-empirical model, motivated by 
the halo occupation framework. Essentially, we assume galaxies obey 
observational constraints on disk masses and gas fractions, and then 
predict the properties of merger remnants. 

We follow a Monte Carlo population of halos from cosmological simulations, 
with merger/growth trees from \citet{fakhouri:halo.merger.rates}. We populate 
every halo with a galaxy\footnote{Starting at some minimum $M_{\rm halo}\sim10^{10}\,\msun$, 
although this choice makes little difference.} on the empirically constrained 
stellar mass-halo mass relation from \citet{conroy:hod.vs.z}. 
As the halo grows, the galaxy is assigned a star formation rate 
and net gas accretion rate (inflow minus outflow) corresponding to 
the track that would be obtained by strictly enforcing that galaxies 
at all times live on the same stellar mass-halo mass relation 
\citep{conroy:hod.vs.z} and gas mass-stellar mass 
relation \citep[fitted from observations as a function of mass 
and redshift in][]{stewart:disk.survival.vs.mergerrates}. 
In other words, 
using the empirical fits to $M_{\rm gal}(M_{\rm halo})$ 
and $f_{\rm gas}(M_{\rm gal})$, we
force galaxies to live on these relations at all times
assuming gas is gained or lost, and stars are formed, exactly as
needed to stay on the empirically fitted correlations, but without a
prior on how much of each is in a bulge or disk component. 
This amounts to the same prescription as assuming galaxies live on 
e.g.\ the observed stellar mass-star formation rate relation 
\citep{noeske:sfh} and tuning accretion rates to reproduce 
the observed dependence of gas fractions on 
stellar mass and redshift \citep{belldejong:disk.sfh,mcgaugh:tf,
calura:sdss.gas.fracs,shapley:z1.abundances,erb:lbg.gasmasses}. 
Observational constraints are present up to redshifts $z\sim3$, 
more than sufficient for our low-redshift predictions.
Despite its simplicity, this effectively guarantees a match to 
various halo occupation statistics including stellar 
mass functions and the fraction of active/passive galaxies 
as a function of mass \citep{yang:obs.clf,weinmann:obs.hod,
weinmann:group.cat.vs.sam,gerke:blue.frac.evol}. 

Together, these simple assumptions are sufficient to define a
``background'' galaxy population. There are degeneracies in 
the model -- however, we are not claiming that this is unique nor that 
it contains any physics thus far. For our
purposes, the precise construction of the empirical model is not
important -- our results are unchanged so long as the same gas
fraction distributions are reproduced as a function of galaxy mass and
redshift.

We illustrate this by comparison with a semi-analytic galaxy formation 
model \citep[][hereafter S08]{somerville:new.sam}. 
A summary of the salient model features is presented in 
Appendix~\ref{sec:appendix:S08}. Unlike the semi-empirical model 
above, the semi-analytic model attempts to predict the relevant 
quantities (e.g.\ disk masses and gas content) from first principles, 
given physical prescriptions for cooling, star formation, and feedback. 
Specifically, the model presented begins with similar halo merger 
trees, allowing for subhalo evolution and stripping, 
and distinguishes between accretion in the ``cold mode'' in 
sub-$L_{\ast}$ galaxy halos (where gas accreted into the halo reaches a 
galaxy on a dynamical time) and ``hot mode'' in 
massive halos (where the cooling occurs from a hot atmosphere, 
suppressed by AGN feedback). The model adopts typical 
recipes for star formation and supernova feedback, motivated 
by observations. 

The relevant physics are entirely contained in the
treatment of mergers. When a merger\footnote{We include all mergers above a 
  minimum mass ratio $\mu\sim0.01$, although our results are not 
  sensitive to this limit, so long as it is small.} 
occurs, we adopt a model for how much of
the gas is drained of angular momentum and participates in a nuclear
starburst, and how much of the stellar disk is violently relaxed ---
both components are removed from the disk component and added to the
mass of the galaxy bulge. If orbital parameters --- namely the
relative inclination angles of the merging disks --- are required, we
simply draw them at random assuming an isotropic distribution of
inclinations (allowing some moderate inclination bias makes no difference). 
We consider three models: \\ \\
{\bf (1) Gas Fraction-Dependent (``Full'') Model:} The prescriptions from
\disksurv, based on a detailed physical model for how disks are
destroyed in mergers and calibrated to the results of 
hydrodynamic merger simulations. In a merger of 
mass ratio $\mu$, a fraction $\sim\mu$ 
of the primary stellar disk is violently relaxed, and a fraction
$\sim(1-\fgas)\,\mu$ of the gas loses its angular momentum. 
For the full details of these scalings as a function of 
mass ratio, orbital parameters, and gas fraction, we
refer to \disksurv, but they are briefly listed in the
Appendix~\ref{sec:appendix}; the general sense is that angular 
momentum loss is suppressed by a
factor $\sim(1-\fgas)$. \\ \\ 
{\bf (2) Gas Fraction-Independent (``Dissipationless'') Model:} The same,
but without accounting for the different behavior of gas and stars
in the merger. We adopt the scaling from \disksurv, but calculate the
angular momentum loss in the gas as if the disks were entirely stellar
--- equivalently, we ignore the fact that gas-richness changes the
efficiency of bulge formation in mergers. In order for gas to
behave in this manner it would have to be dissipationless and
collisionless; but nevertheless this is a common (almost ubiquitous)
assumption. The specific scalings are given in Appendix~\ref{sec:appendix}. \\ \\ 
{\bf (3) Simplified (``Threshold'') Model:} A ``threshold'' (``major''
versus ``minor'') assumption, common in many semi-analytic models. 
In major mergers ($\mu>0.3$), the model assumes that the entire 
stellar disk is violently relaxed and the entire cold gas content 
participates in a burst. In minor mergers, the primary is unperturbed, the 
stars of the secondary are added to the bulge, and the secondary 
gas content is added to the disk. \\

\section{Consequences}
\label{sec:consequences}

\subsection{Bulges and Disks at $z=0$}
\label{sec:BT}

\begin{figure*}
    \centering
    \scaleup
    \plotside{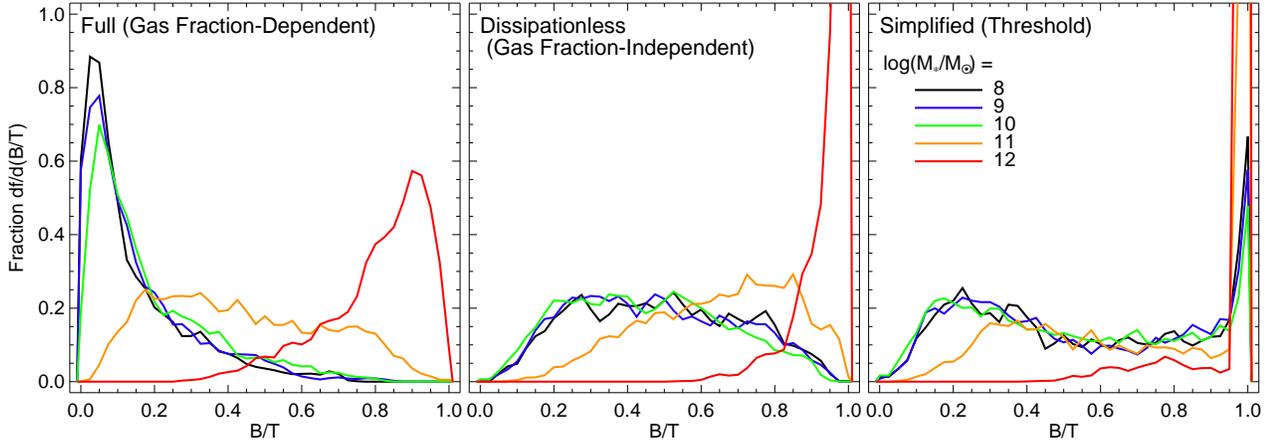}
    \caption{Predicted distribution of bulge-to-total stellar mass
      ratios in galaxies of
      different stellar masses at $z=0$, from our 
      semi-empirical model, with different models for 
      bulge formation in mergers (see text). 
      {\em Left:} The full (gas fraction-dependent) 
      model from \disksurv: bulge formation is suppressed in gas-rich 
      mergers. {\em Center:} The ``dissipationless'' model: 
      bulge formation is gas fraction-independent. 
      {\em Right:} A simplified ``threshold'' model, also 
      gas fraction-independent, with simplified treatment of the 
      dependence on merger mass ratio. 
      Accounting for the gas fraction-dependent physics of 
      angular momentum loss in mergers, 
      the predicted $B/T$ is suppressed in low mass, gas-rich galaxies. 
    \label{fig:BT.pred}}
\end{figure*}

\begin{figure}
    \centering
    \scaleup
    \plotone{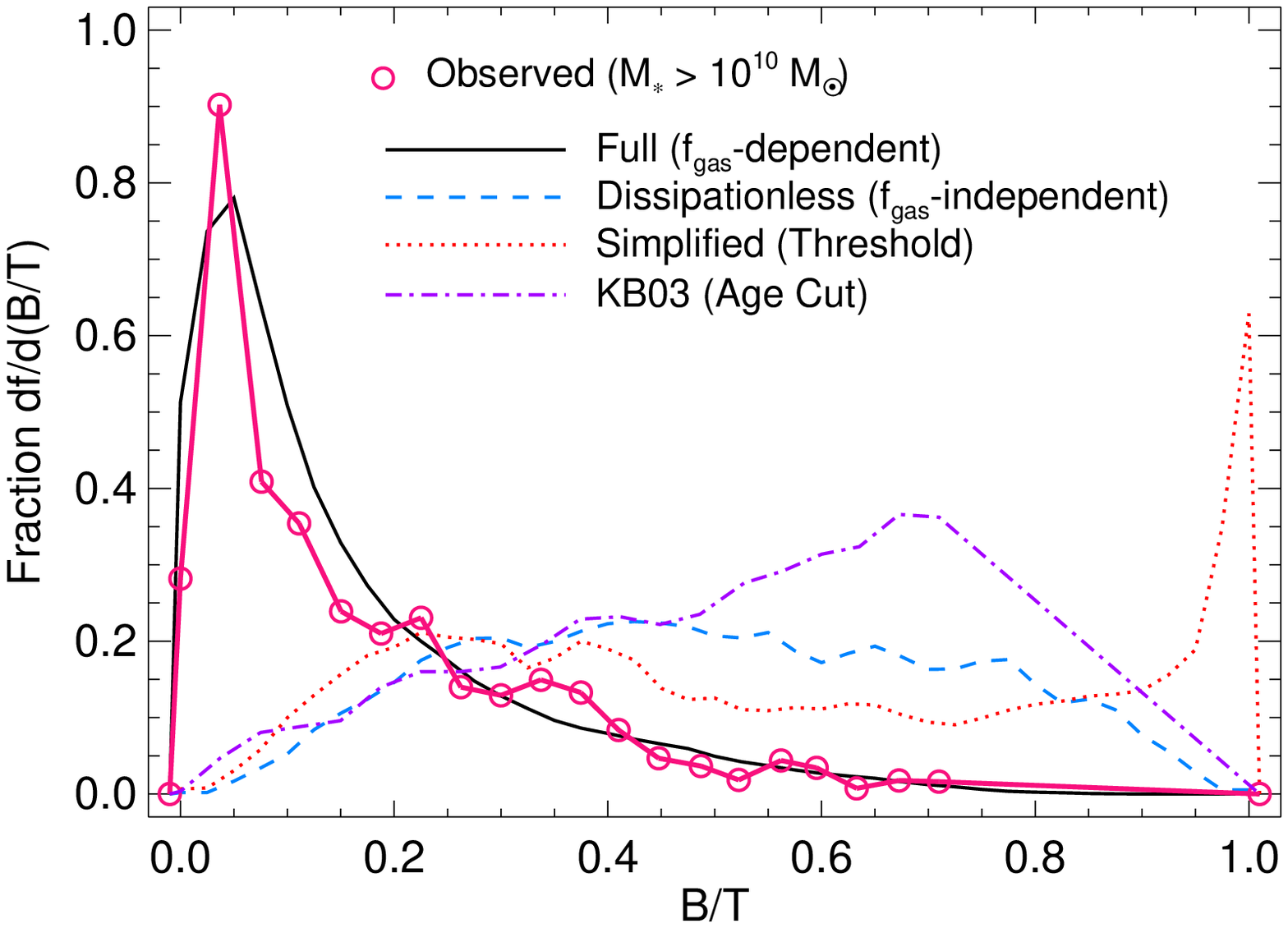}
    \caption{As Figure~\ref{fig:BT.pred}, for galaxies with $M_{\ast}>10^{10}\,\msun$. 
    We compare the models from Figure~\ref{fig:BT.pred} with observations 
    from \citet{weinzirl:b.t.dist}. We also compare the semi-analytic 
    model from \citet{khochfar:sam}, which adopts different assumptions and 
    does not allow mergers at $z>2$ to form bulges, but employs 
    the simplified (threshold) model for bulge formation at $z<2$. 
    Only the gas fraction-dependent model matches the observations; 
    suppressing high-$z$ bulges alone is insufficient to resolve the discrepancy 
    in gas fraction-independent models. 
    \label{fig:BT.vs.obs}}
\end{figure}

Figure~\ref{fig:BT.pred} shows the predicted distribution of
bulge-to-total (bulge stellar mass divided by total galaxy stellar
mass) ratios as a function of galaxy mass. 
Figure~\ref{fig:BT.vs.obs} compares these predictions 
to observations of field spirals from \citet{weinzirl:b.t.dist}.\footnote{The 
  observed $B/T$ are determined in $H$-band; the models in 
  Figure~\ref{fig:BT.vs.obs} predict $B/T$ in terms of stellar mass. 
  In the semi-analytic model, the predicted star formation histories 
  are used to predict the $H$-band $B/T$; generally 
  we find this is slightly smaller than the stellar mass $B/T$, 
  owing to the lower mass-to-light ratio in the younger stellar disk.}
At high masses, all three models predict most galaxies are bulge-dominated. 
At low ($<10^{10}\,\msun$) and
intermediate ($\sim10^{10}-10^{11}\,\msun$) masses, there is a
dramatic difference: when the role of gas in suppressing disk
destruction in mergers is accounted for, the model predicts the
existence of considerably more disk-dominated systems. 

The results are very similar when we adopt the semi-analytic approach
from S08.  We also compare the semi-analytic model
of \citet{khochfar:sam}, which relies on different cosmological
assumptions and disk formation models, but still
assumes the ``threshold''-like (gas fraction-independent)
model for bulge formation in mergers, leading to 
a too bulge-dominated population. The authors attempt to address this 
with an age cut: enforcing that all mergers at $z>2$ make no bulge. 
Figure~\ref{fig:BT.vs.obs} shows this is insufficient: many of these systems 
have too many mergers at late times to invoke redshift-dependent dynamics 
or rapid subsequent cooling to ``offset'' the bulge formed.

\begin{figure}
    \centering
    \scaleup
    \plotone{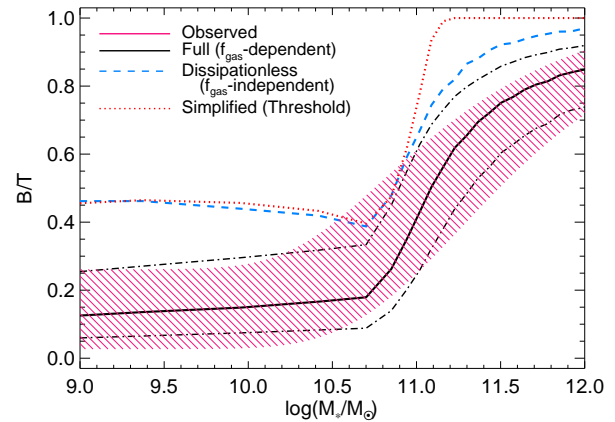}
    \caption{Mean $B/T$ as a function of galaxy stellar 
    mass in observations and our semi-empirical models. 
    Dot-dashed black lines show the $\pm1\,\sigma$ scatter 
    in the full (gas fraction-dependent) model. The shaded magenta 
    range is the observed relation (with scatter) from 
    \citet{balcells:bulge.scaling}. The models are all similar
    at $>L^{\ast}$ masses/luminosities, where disks are relatively
    gas-poor; at lower mass, the differences from
    Figures~\ref{fig:BT.pred}-\ref{fig:BT.vs.obs} are apparent. 
    \label{fig:BT.vs.m}}
\end{figure}

Figure~\ref{fig:BT.vs.m} compares the
median $B/T$ (and scatter)
as a function of stellar mass for each model to that observed. 
The predicted scatter owes mostly to differences in merger history. 
We show the fits to observations in \citet{balcells:bulge.scaling}, but others are similar
\citep{kochanek:morph.stellar.mf,allen:bulge-disk,
noordermeer:bulge-disk,driver:bulge.mfs}. 
All the models predict some mass dependence, as merger 
histories are not completely mass-independent. However, at low
masses, the gas fraction-independent 
models predict $B/T\sim0.4-0.5$ (E/S0 galaxies), much larger 
than observed. 
Note that both gas fraction-independent models give similar results; 
tuning the relative importance of major/minor mergers does not 
resolve their problems. 

\begin{figure}
    \centering
    \scaleup
    \plotonesmallish{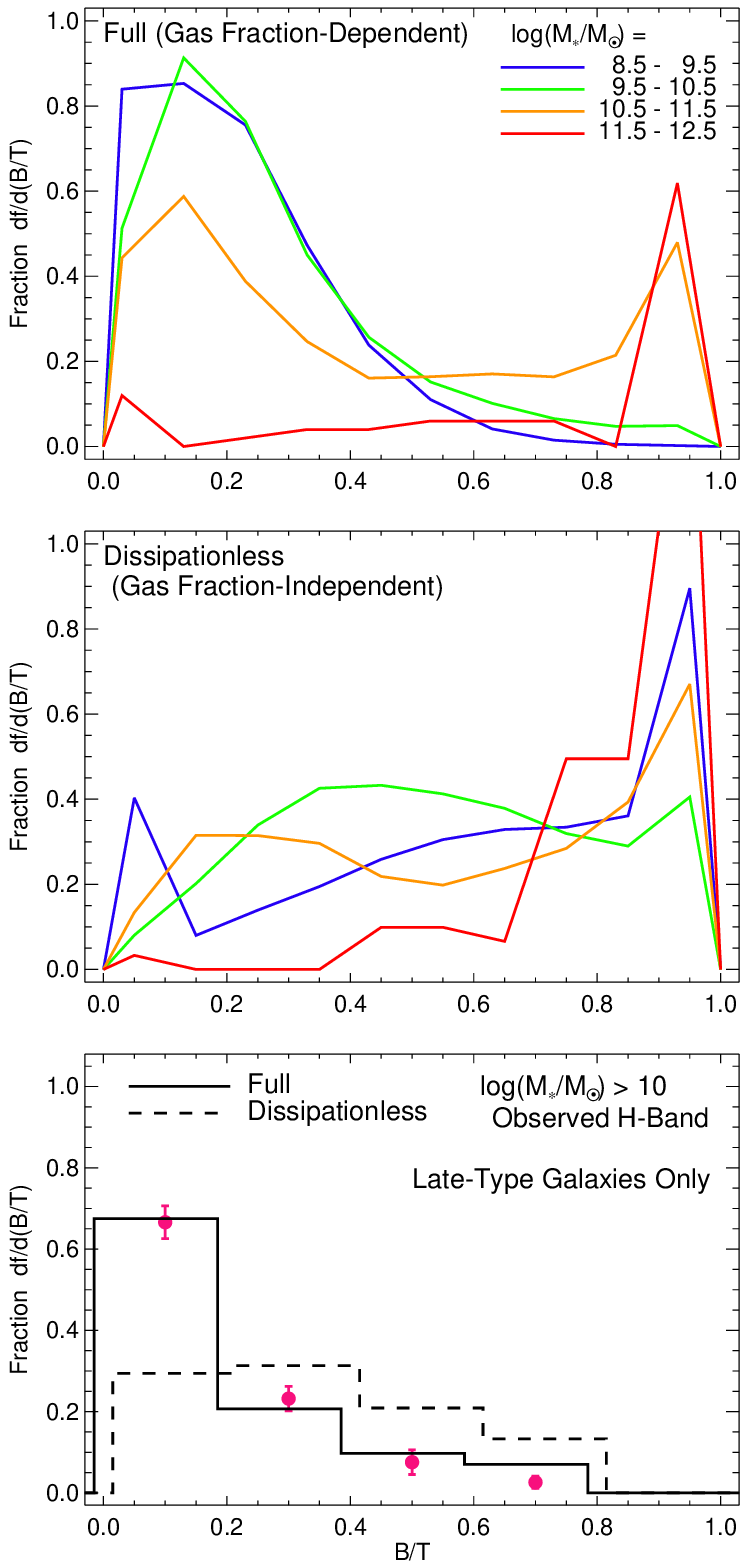}
    \caption{As Figures~\ref{fig:BT.pred} \&\ \ref{fig:BT.vs.obs}; but
      showing the results of the semi-analytic model, with either the
      full (gas fraction-dependent) bulge formation model or
      threshold (gas fraction-independent) model. 
      {\em Bottom:} The results for
      $M_{\ast}>10^{10}\,\msun$ galaxies with $B/T$ computed in the
      $H$-band, pre-selected to include {\em only} late-type galaxies
      ($B/T<0.75$), compared to the observations of \citet{weinzirl:b.t.dist}. 
      The semi-analytic and semi-empirical models yield similar conclusions. 
    \label{fig:BT.vs.obs.rachel}}
\end{figure}

Figure~\ref{fig:BT.vs.obs.rachel} shows the results from the
semi-analytic model for the same quantities considered in
Figures~\ref{fig:BT.pred} \&\ \ref{fig:BT.vs.obs}. 
The differences between the predicted bulge-to-total distributions are
dominated by the different treatment of the role of gas in
mergers, rather than by the methodologies or different choices in the
galaxy formation model.


\begin{figure*}
    \centering
    \scaleup
    \plotside{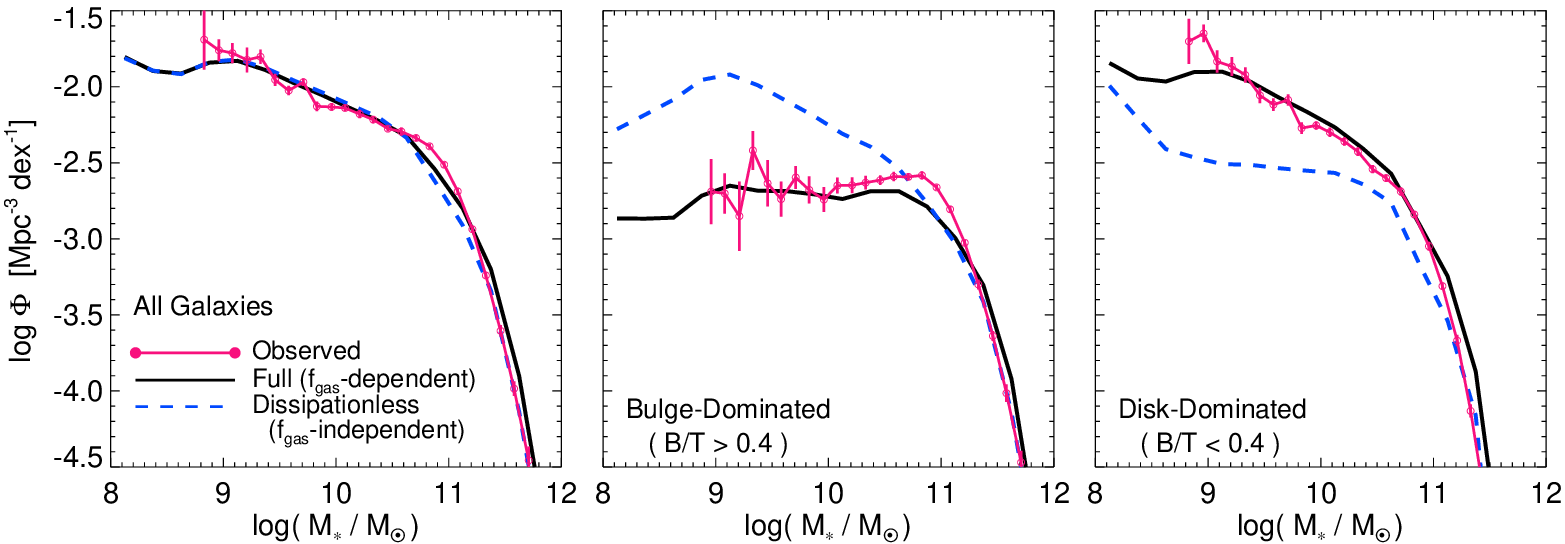}
    \caption{Comparison of the $z=0$ mass
      function of all ({\em left}), bulge-dominated ({\em center}),
      and disk-dominated ({\em right}) galaxies.  We compare the
      gas fraction-independent (threshold) and gas
      fraction-dependent (full) model, in the
      semi-analytic model, to the observed morphology-dependent mass
      functions from \citet{bell:mfs}.  Despite
      freedom in the SAM to adjust prescriptions, gas
      fraction-independent models overpredict the low-mass
      bulge-dominated population. Including
      the physical dependence on gas fraction reconciles
      this discrepancy.
    \label{fig:rachel.mf}}
\end{figure*}

\begin{figure}
    \centering
    \scaleup
    \plotone{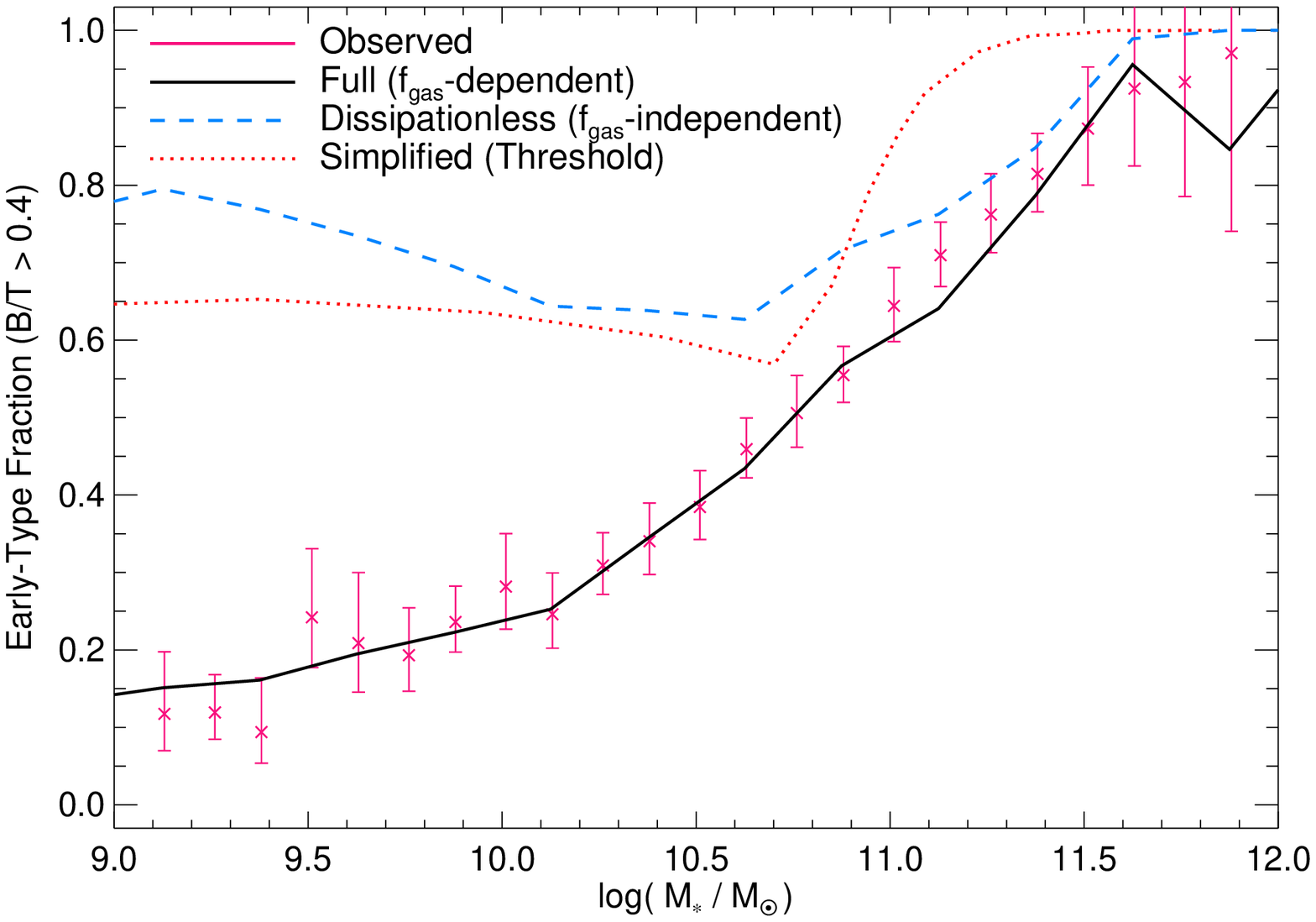}
    \caption{Early-type ($B/T>0.4$) fraction as a
      function of stellar mass. We compare
      with the $z=0$ observed trend from the morphologically
      classified mass functions in \citet{bell:mfs}.
    \label{fig:ellfrac}}
\end{figure}

Figure~\ref{fig:rachel.mf} compares the predicted stellar mass functions 
in the semi-analytic model. 
Figure~\ref{fig:ellfrac} shows the  corresponding 
early-type fraction as a function of stellar mass, 
compared with the results from the morphologically
classified stellar mass functions of \citet{bell:mfs} \citep[see
also][who obtain similar results]{kochanek:morph.stellar.mf,
bundy:mfs,pannella:mfs,franceschini:mfs}. 
Although the gas-fraction independent 
models yield good agreement with the {\em total} stellar mass 
function, they predict an excess of bulge-dominated galaxies 
at low masses. This is robust to the exact 
morphology or $B/T$ cut used to define bulge or disk-dominated
systems; here we adopt $B/T\sim0.4$ as a dividing line as this is a
good proxy, in simulations, for the concentration cut in
\citet{bell:mfs}. 

We emphasize that we have incorporated the bulge formation
prescriptions directly from \disksurv\, with the values of all related
parameters taken directly from the results of the hydrodynamic
simulations. We have not changed any of the other physical recipes in
the SAM, nor have we re-tuned any of the baseline free parameters. We
find it particularly convincing that the gas-fraction dependent
bulge-formation prescription based on the simulations immediately
produced excellent agreement with the observations at all mass scales,
with no fine-tuning.

\subsection{Redshift Evolution}
\label{sec:z}

\begin{figure}
    \centering
    \scaleup
    \plotone{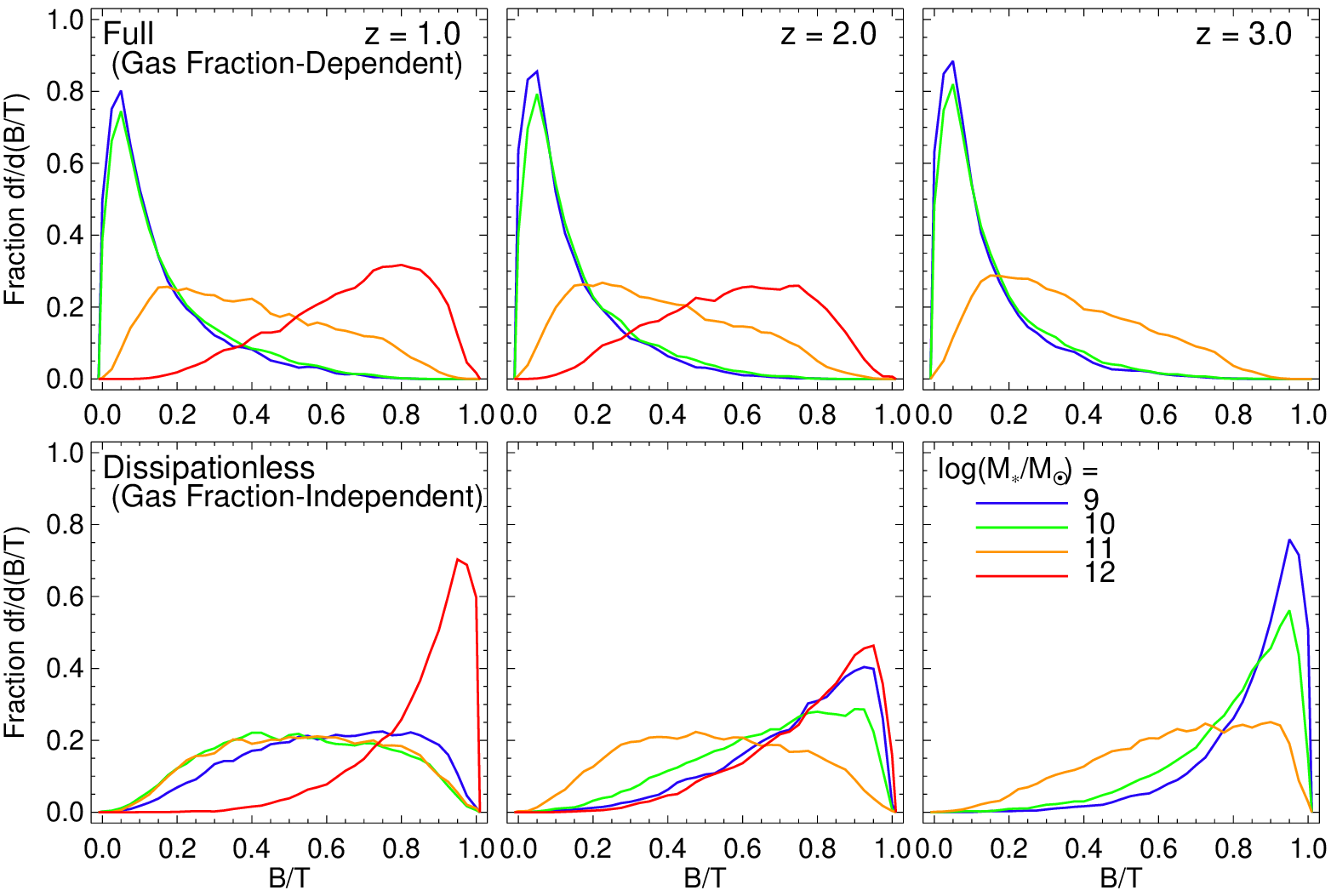}
    \caption{As Figure~\ref{fig:BT.pred}, but evaluated at different
      redshifts $z=1$, $z=2$, $z=3$, for our gas
      fraction-dependent model ({\em top}) and gas
      fraction-independent model ({\em bottom}), implemented within
      the semi-empirical framework.
      The gas fraction-independent model predicts that the
      population is {\em more}
      bulge-dominated at higher redshifts (because merger rates are
      higher). Including the gas dependence reverses this:
      bulges are less dominant at high redshifts, as
      gas fractions are systematically higher.
    \label{fig:BT.dist.z}}
\end{figure}

\begin{figure}
    \centering
    \scaleup
    \plotonesmall{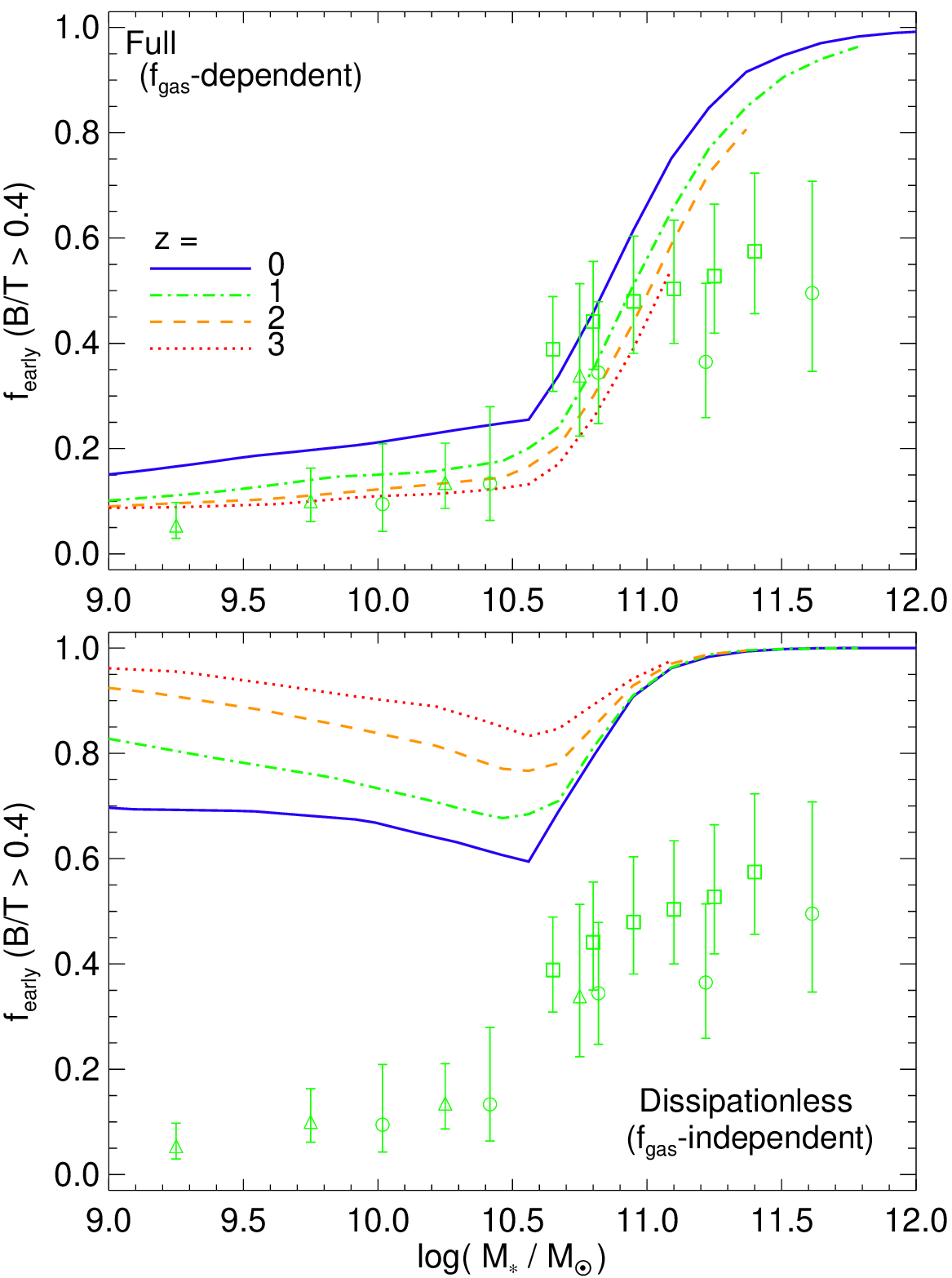}
    \caption{Early-type fraction at different
      redshifts, in the semi-empirical model. 
      Green points show the observations from
      \citet[][circles]{bundy:mfs}, \citet[][squares]{ilbert:cosmos.morph.mfs}, and 
      \citet[][triangles]{pannella:mfs} from the 
      COSMOS, DEEP and GOODS fields
      at $z\approx1$ (compare the green dot-dashed
      line). Ignoring the gas-fraction dependence predicts that even low-mass
      populations are $\gtrsim 80\%$ elliptical at $z>1$.
    \label{fig:morph.mass.evol}}
\end{figure}

Figure~\ref{fig:BT.dist.z} repeats the 
comparison from Figure~\ref{fig:BT.pred} at $z=1,\,2,\,3$. 
Figure~\ref{fig:morph.mass.evol} plots the median $B/T$ and early-type 
fraction versus redshift and stellar mass. In the gas fraction-dependent 
model, higher gas fractions lead to more disk-dominated 
galaxies at high redshifts (see \S~\ref{sec:how}). 
The evolution in gas 
fractions that we adopt \citep[fitted to the observations in][]{erb:lbg.gasmasses} 
is relatively weak; if real evolution in gas fractions is stronger, it will 
further suppress $B/T$. 

Merger rates were higher in the
past, so the gas fraction-independent 
models predict that galaxies should be {\em more} bulge-dominated. By
$z\sim1-2$, such models would predict that essentially all galaxies at
all masses are bulge-dominated.

At the highest redshifts, morphological determinations are difficult; however, 
there are some measurements at $z\sim1$. 
We compare to the results from
\citet{bundy:mfs} from the GOODS fields, \citet{pannella:mfs} from
the GOODS South and FORS Deep fields, 
and \citet{ilbert:cosmos.morph.mfs} from the COSMOS fields. 
Despite large uncertainties, the trend in the fraction of
bulge-dominated galaxies with mass is qualitatively very similar to
that observed at $z=0$, but with a systematically 
lower bulge-dominated (E/S0) fraction at all
masses.

This agrees well with the expectations from the gas fraction-dependent 
model, and strongly conflicts with the gas
fraction-independent model prediction. If anything, the data suggest
that at high masses, bulges may be even more suppressed than the
dissipational model prediction; as noted above, this could be
explained if gas fractions at these masses evolve more rapidly. 
Theoretical models suggest, for example, that the efficient cooling 
(``cold mode'') regime may persist to high masses at high 
redshifts \citep{keres:hot.halos,keres:cooling.revised,dekel:cold.streams}.
However, cosmic variance (not included in the quoted errors) 
remains a significant source of uncertainty at high masses.

\begin{figure}
    \centering
    \scaleup
    \plotone{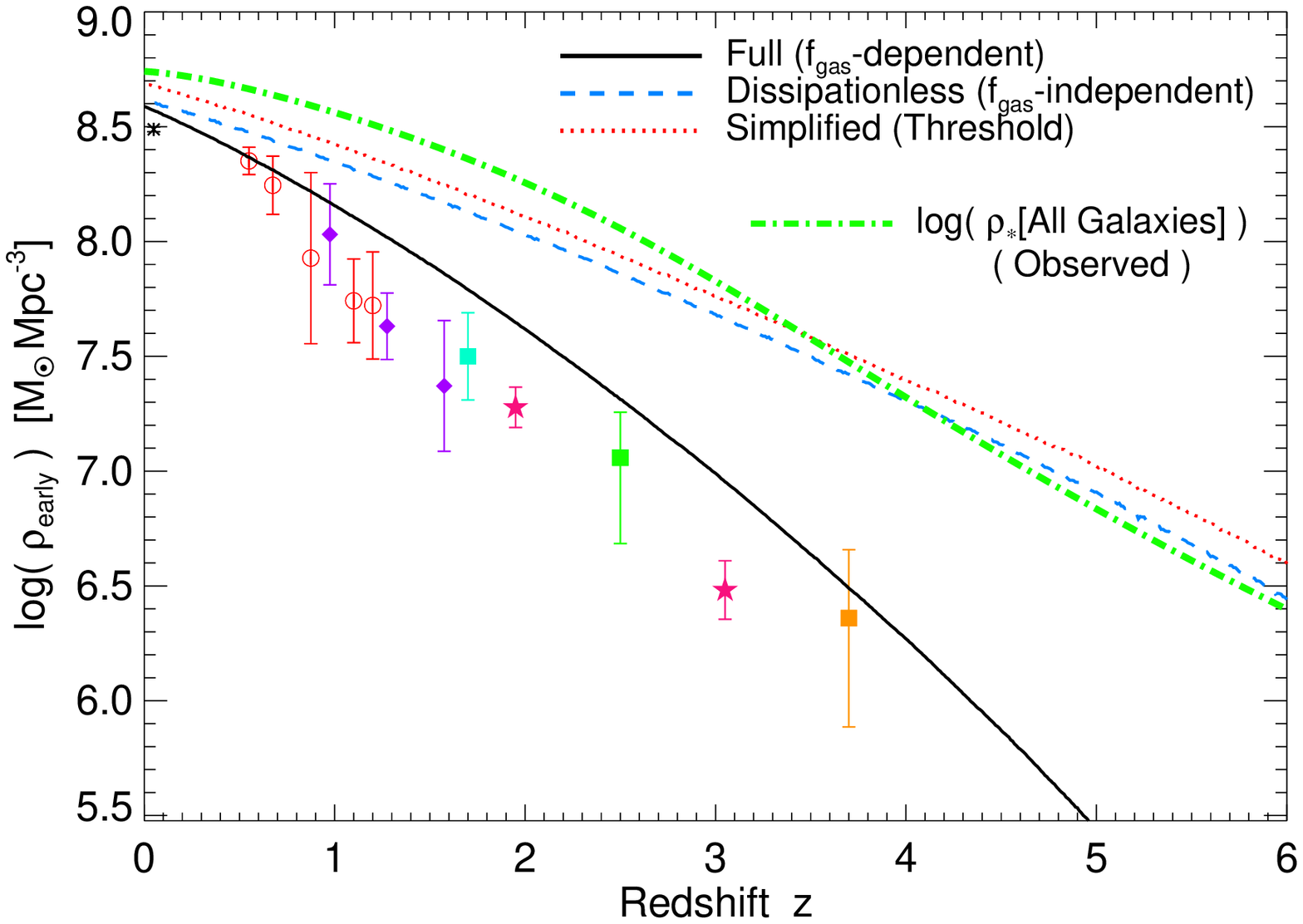}
    \caption{Integrated mass density in bulge-dominated
      galaxies in the semi-empirical models as a function of redshift, 
      compared to observations (points). 
      Contrast the observed total stellar mass density 
      \citep[][dot-dashed line]{hopkinsbeacom:sfh}. 
      Observations are compiled from the 
      morphologically-selected samples of \citet[][black
        $\times$]{bell:mfs}, \citet[][red
        circles]{bundy:mfs,bundy:mtrans}, \citet[][violet
        diamonds]{abraham:red.mass.density}, and \citet[][cyan
        square]{daddi05:drgs}, and color-selected samples of
      \citet[][green square]{labbe05:drgs}, \citet[][orange
        square]{vandokkum06:drgs}, and \citet[][magenta
        stars]{grazian:drg.comparisons}.
    \label{fig:rhobul.evol}}
\end{figure}

Figure~\ref{fig:rhobul.evol} compares the redshift evolution of the
predicted bulge mass density\footnote{Specifically,
  we plot the mass density in bulge-dominated galaxies,
  which is not the same as the absolute mass density in all bulges,
  but is closer to the observed quantity. At high redshifts 
  $z>1.5$ observed morphologies are ambiguous; we show 
  the mass density in passively
  evolving red galaxies as a proxy. This may not be appropriate, 
  but at $z<1$ the two correspond well, and 
  the compactness, size, and kinematics of the ``passive'' objects 
  do appear distinct from star-forming objects 
  \citep{kriek:drg.seds,toft:z2.sizes.vs.sfr,trujillo:ell.size.evol.update,
  franx:size.evol,genzel:highz.rapid.secular}.}  
As expected from
Figure~\ref{fig:morph.mass.evol}, in the gas
fraction-independent models, 
the predicted bulge mass density approaches the global stellar mass
density at high-$z$.  In gas fraction-dependent 
models, the trend is reversed: at high redshift, higher gas fractions
suppress bulge formation. The latter trend agrees
with the available observational results.  

Though important for bulges, 
we find that these differences have little effect on the 
global star formation rate density. 
``Quiescent'' (non merger-driven) star formation dominates 
at all redshifts \citep[see also][]{wolf:merger.mf,
  menanteau:morphology.vs.sfr,hopkins:merger.lfs,
  noeske:sfh,jogee:merger.density.08}. Differences 
owing to e.g.\ systems ``retaining'' more gas for later consumption 
are much smaller than the uncertainties from e.g.\ star formation rate and 
stellar wind prescriptions.

\subsection{Why Does It Work?}
\label{sec:how}

\begin{figure}
    \centering
    \scaleup
    \plotone{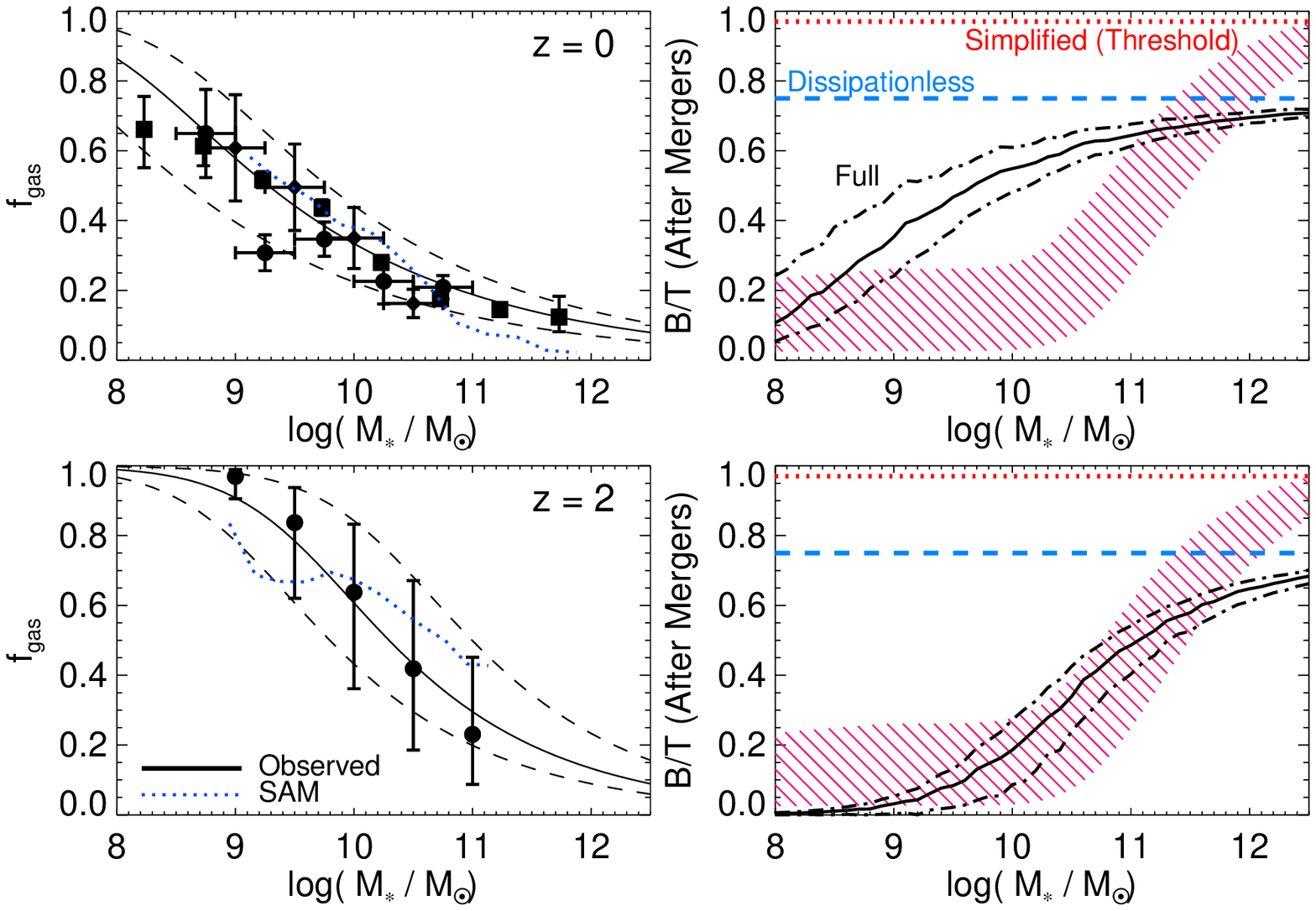}
    \caption{{\em Top Left:}
      Observed median gas fraction as a function of stellar mass in
      $z=0$ disks, from \citet[][diamonds]{belldejong:tf}, 
      \citet[][squares]{kannappan:gfs}, and \citet[][circles]{mcgaugh:tf}, with an 
      analytic fit (solid line) and $\pm1\,\sigma$ scatter (dashed). 
      Dotted (blue) line compares median gas fractions in
      the semi-analytic model.  
      {\em Top Right:} Predicted 
      bulge-to-total ratio in the remnants if every such
      galaxy immediately underwent a series of mergers that doubled
      its stellar mass (style as Figure~\ref{fig:BT.vs.m}). 
      Lines correspond to the different models: full/gas fraction-dependent 
      (solid black), dissipationless/gas fraction-independent (dashed 
      blue), and simplified/threshold (dotted red). 
      Compare the observed trend (shaded). 
      {\em Bottom:} Same, but based on
      the gas fractions at $z\sim2$
      \citep{erb:lbg.gasmasses}. Accounting for the full model results 
      from simulations, high gas fractions in low-mass systems suppress 
      bulge formation. 
    \label{fig:fgas.fbulge}}
\end{figure}

Why does the revised bulge formation model produce the observed
trends?  Figure~\ref{fig:fgas.fbulge} illustrates the relevant
physics.  Consider the observed gas fractions of disks as a function
of stellar mass, shown in the Figure to be a strong decreasing
function of mass.\footnote{At $z=0$, the gas fractions shown are based
  on measured atomic HI gas fractions; \citet{belldejong:tf} correct
  this to include both He and molecular H$_{2}$; \citet{mcgaugh:tf}
  correct for He but not H$_{2}$; \citet{kannappan:gfs} gives just the
  atomic HI gas fractions \citep[this leads to slightly lower
    estimates, but still within the range of uncertainty plotted;
    H$_{2}$ may account for $\sim20-30\%$ of the dynamical mass, per
    the measurements in][]{jogee:H2.masses}.  We emphasize that these
  gas fractions are lower limits (based on observed HI flux in some
  annulus).  At $z=2$,
  direct measurements are not available; the gas masses from
  \citet{erb:lbg.gasmasses} are estimated indirectly based on the
  observed surface densities of star formation and assuming that the
  $z=0$ Kennicutt law holds; other indirect estimates 
  yield similar results \citep{cresci:dynamics.highz.disks}.} 
Now assume that every such
galaxy doubles its mass in a merger or sequence of
mergers\footnote{This could be a single 1:1 merger,
  three 1:3 mergers, or even ten 1:10 mergers; because the
  efficiency of bulge formation as calibrated in simulations is nearly
  linear in mass ratio, they give indistinguishable results for our
  purposes here, so long as they
  all occur within a short timescale \citep[see
    e.g.][]{naab:minor.mergers,bournaud:minor.mergers,
    cox:massratio.starbursts,hopkins:disk.survival}. 
  In practice, this is unlikely at
  smaller mass ratios; cosmological calculations show that
  e.g.\ ten 1:10 mergers are a less likely channel for growth of
  spheroids than three 1:3 mergers \citep[][]{gottlober:merger.rate.vs.env,
    maller:sph.merger.rates,zheng:hod.evolution,stewart:mw.minor.accretion}.},
and consider the resulting $B/T$ at each mass. This represents a 
major perturbation, and in any of the gas fraction-independent models would 
uniformly transform every galaxy into nearly $100\%$ bulge, 
independent of mass. 

As discussed in \S~\ref{sec:intro}, however, gas loses angular momentum 
to internal torques from the stellar disk. If the disk is mostly 
gas, then the stellar bar/perturbation that does the majority of the work 
torquing the gas is relatively less massive and so less gas 
falls to the center and starbursts. In a collision of two pure gas 
disks, there is no stellar feature to torque the gas 
(in fact, accounting for typical orbital angular momenta, such a collision 
typically increases the disk specific angular momentum). 

Given these physics and the observed trend of gas fraction with 
mass, we expect low mass 
galaxies ($M_{\ast}\lesssim10^{9}\,\msun$) to have just $\sim20\%$ 
bulge even after a recent 1:1 merger. If mergers occur at higher 
redshift, when gas fractions are higher, the suppression is 
more efficient. It is possible for every 
galaxy to double its mass at $z\sim1-3$ in major mergers and still lie below 
the local $B/T$-mass correlation (allowing some later mergers). In practice, 
many of these bulges 
are made by minor mergers, but the relative suppression is similar 
\citep[][]{weinzirl:b.t.dist,hopkins:merger.rates}. 

This accounts for most of the observed trend in bulge fraction versus 
stellar mass. However, it is supplemented by two other trends 
(the origin of the weak scaling in even the 
gas fraction-independent models): more massive galaxies tend to have 
had more violent merger histories, and to have lower cooling rates
(leading to less disk re-growth and lower gas fractions).  

\begin{figure}
    \centering
    \scaleup
    \plotonesmallish{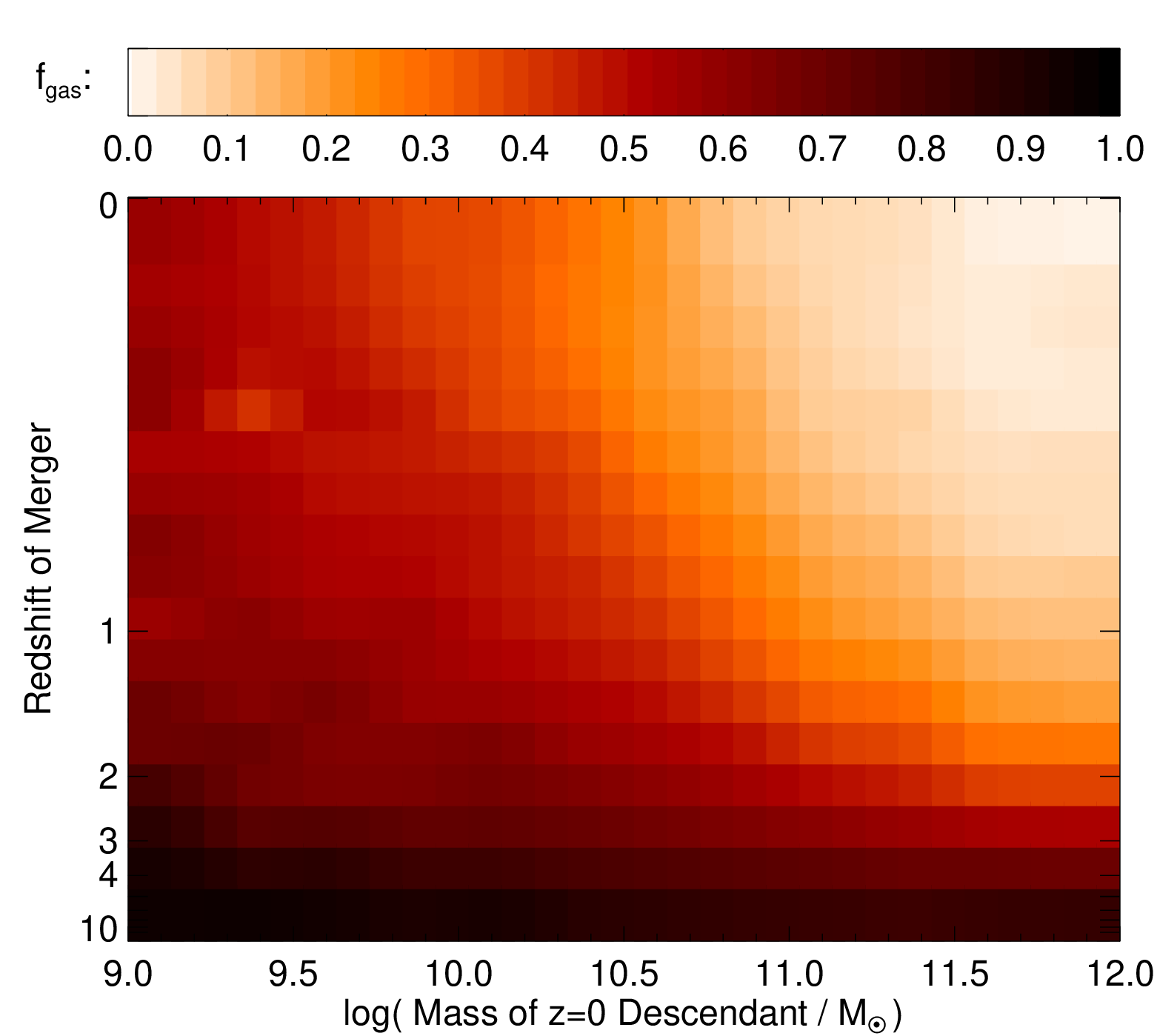}
    \caption{Galaxy gas fractions as a function of the $z=0$ stellar
      mass of the descendant and the time at which the merger
      occurred (in the semi-analytic model). 
      Specifically, considering galaxies which have the given
      $z=0$ stellar mass, and whose main branch progenitor experienced
      a merger at the given redshift (scaled here in equal-cosmic time
      units), we plot the median gas fraction of the progenitors just
      before the merger(s).  All galaxies were more gas-rich at high
      redshift, but low-mass galaxies are more gas-rich than
      high mass galaxies (on average) at all epochs. 
    \label{fig:fgas.z}}
\end{figure}

Figure~\ref{fig:fgas.z} illustrates this in another way --- we show
the distribution of gas fractions in mergers as a function of merger 
redshift and stellar mass of the $z=0$ descendant galaxy ({\em not}
instantaneous mass at the time of the merger). Gas fractions
decrease with time overall, but low-mass galaxies are always more
gas-rich than high-mass galaxies. This is for two reasons: (1) the
Kennicutt law predicts that star formation, and therefore gas
consumption, is more efficient in high mass, higher surface density 
galaxies; (2) cooling and accretion are quenched in high-mass halos 
(in semi-analytic models, this is typically attributed to heating of ``hot'' quasi-static 
halos by AGN feedback). 

Figures~\ref{fig:fgas.fbulge}
\&\ \ref{fig:fgas.z} illustrate that the dividing line between
gas-rich and gas-poor galaxies moves to higher masses at higher
redshifts, suppressing bulge
formation even in massive galaxies at high redshifts. 
In semi-analytic models, this evolution occurs for three
reasons: (1) cooling rates increase uniformly with redshift, (2) the
critical halo mass dividing the ``cold'' and ``hot'' accretion modes
(below which accretion is efficient) likewise increases with
increasing cosmic gas densities at higher redshifts, and (3) AGN
feedback becomes less efficient at high redshifts. 
For most of the $z=0$ population, there is a
relatively narrow ``window'' of efficient bulge formation, where
cooling slows down, 
galaxies consume most of their gas, 
and mergers become efficient at forming bulges.

Disk survival and bulge formation are at some level
zero-sum. However, because various processes contribute to disk growth
at different times, it is useful to ask whether disk ``survival'' or the suppression 
of bulge growth is more important. In
order to answer this, we consider the following simple experiment: in
each merger, we destroy as much of the disk gas and stellar mass as
predicted by the dissipationless (gas fraction-independent) model;
however, we only add to the bulge the mass that the dissipational (gas
fraction-dependent) model says should be violently relaxed or lose
angular momentum. The remaining mass is thrown away. This is not
physical, but it allows us to see whether our conclusions
would change if bulge formation was suppressed per our full
gas fraction-dependent model but disks were still relatively efficiently
destroyed in mergers (per our gas fraction-independent
model).

We find that the results 
are almost unchanged from the full
gas fraction-dependent model. 
The over-prediction in gas fraction-independent models of $B/T$ 
is a problem of too much bulge formation, rather than too little 
disk survival from early times. This should not be surprising; various 
constraints suggest that low-mass galaxies are growing rapidly, and 
so disk masses are usually dominated by recent accretion and 
star formation \citep[see e.g.][and references therein]{belldejong:disk.sfh,
gallazzi:ssps,noeske:sfh}.

\section{Discussion}
\label{sec:discuss}

A realistic model for how gas loses angular
momentum and builds bulges in mergers qualitatively
changes the expectations for the global distribution of bulge to disk
ratios in the local Universe and the efficiency of bulge formation as
a function of mass and redshift. 
As demonstrated in simulations in \disksurv, 
gas in mergers loses its angular
momentum to {\em internal} torques from the induced non-axisymmetry in
the stellar distribution in the same disk as the gas. Therefore,
in the limit of very gas rich mergers (with little or no stellar mass
in the disks to do such torquing), angular momentum loss
and destruction of the gas disk is inefficient (to lowest order, the
efficiency scales $\propto (1-\fgas)$).

We incorporate this idea into a cosmological framework using two
different approaches: a simple semi-empirical model and a full semi-analytic 
model. The conclusions appear robust to this choice; we have also 
experimented with a number of variations to the semi-empirical 
model (detailed comparison of the resulting differences for 
merger rates will be the subject of future work; we briefly summarize 
the variations in Appendix~\ref{sec:appendix:variations}), 
and find similar results in each case. \citet{stewart:disk.survival.vs.mergerrates} 
also consider an independent semi-empirical model, using different 
halo merger trees, observational constraints, and more qualitative 
prescriptions for behavior in mergers (rather than the detailed 
simulation results here), and reach the same 
conclusions. It appears that, 
regardless of the methodology, this picture of 
morphological transformation in mergers makes qualitatively different
predictions from the ``traditional'' picture which ignores the
dependence on gas fraction. Because, in the local Universe, lower mass
disks have large gas fractions ($\fgas\gtrsim0.5$ at
$M_{\ast}<10^{10}\,M_{\sun}$), this in particular implies stronger
suppression of bulge formation at low masses.  Even if the
dissipationless (halo) merger histories were self-similar across
different mass scales, this trend of gas fraction with mass breaks
this self-similarity and largely explains the observed trend of bulge
fraction with mass.

Other factors are expected to weakly break this self-similarity (the
non-linear dependence of galaxy baryonic mass on halo mass, and weak
scale dependence in the dark matter merger histories), 
but simulations and empirical constraints have found that accounting 
for these factors alone predicts too weak a trend. 
On the other hand, allowing for the role of gas-richness
(implicitly related to the nature of cooling and star formation
efficiency) provides a natural and predictive
explanation for the observed mass-morphology relation.

Moreover, whereas simulations without large gas supplies, or 
gas fraction-independent models have difficulty 
reproducing the abundance of ``bulge-less'' or 
low $B/T$ disks ($B/T\lesssim0.1-0.2$),
this problem is largely solved by incorporating the gas
fraction-dependent bulge formation prescription. In practice, if
galaxies have sufficiently low star formation efficiencies in 
order to retain very large gas fractions, as implied
by observations of these systems, and especially if the gas distribution is very extended (as
in e.g.\ low surface brightness systems), then it is possible to
produce systems with such low $B/T$ even in the immediate aftermath of
a major merger, let alone in systems which have gone $\sim10$\,Gyr
since the last such merger with significant subsequent accretion. In
practice, many bulges of low (and, to a lesser extent, intermediate)
mass systems are predominantly made by minor mergers, making the
required gas-richness even more moderate.

At low redshifts, this is primarily relevant for small-mass galaxies,
where gas fractions are high. However,
observations indicate that gas fractions 
should be larger at high redshifts, such that even some 
$\sim\lstar$ galaxies appear to have $\fgas\gtrsim0.5$ at redshifts
$z>2-3$. At increasing redshift, therefore, the effects described
above propagate to higher and higher masses, and become important for
an increasingly large proportion of the galaxy population.

This is seen as well in recent cosmological simulations
\citep{governato:disk.formation,governato:disk.rebuilding}: the
presence of gas leads to non-destructive mergers and rapid disk
re-assembly. The role of stellar feedback is to preserve
gas, preventing it from all turning into stars at early times or in
very small galaxies, and to
remove low angular momentum gas from the disk (keeping the gas at
relatively large radii, comparable to observed disk sizes). Even at
$z\sim0.5$, disks experience major mergers in these simulations and
rapidly re-build. These processes, it appears, are now being seen 
in a growing number of observations
\citep{hammer:obs.disks.w.mergers,zheng:morphological.sfr.evol,
  trujillo:truncation.scale.evol,
  flores:tf.evolution,puech:minor.merger.at.z06,
  puech:highz.vsigma.disks,puech:tf.evol,
  atkinson:scatter.in.tf.owes.to.increased.mergers}.  The suppression
of bulge formation in the merger owing to its gas-richness, combined
with rapid large-scale gas inflows concurrent with and subsequent to
the merger, both contribute to the rapid re-building of a
disk-dominated galaxy following the event
\citep{springel:spiral.in.merger,robertson:disk.formation}.  The
inclusion of this continuous accretion means that a very large gas
fraction $f_{\rm gas} > 0.5$ is not needed to build a disk-dominated
galaxy (as would be the case if the system received no new gas after
the merger); the more moderate bulge suppression that would result
from intermediate gas fractions is sufficient.

At higher redshifts, merger rates increase, but in gas
fraction-dependent models this is more than offset by the increase in
disk gas content and cooling efficiencies, and thus low $B/T$ ratios
are maintained. However, because the rates of these processes (mergers
and cooling/accretion) scale with the Hubble time while the internal
relaxation times of disks are relatively constant, disks should be
more turbulent and the relaxation from the
building/merger/re-formation process should be evident in more
systems, as thicker, more dispersion-dominated disks with enhanced
incidence of structure and clumpiness.  In fact, this is widely
observed, \citep{weiner:disk.kinematics.evol,
  atkinson:scatter.in.tf.owes.to.increased.mergers,
  puech:highz.vsigma.disks,neichel:disks.disturbed.at.higher.z,
  yang:disk.kinematics.unrelaxed.at.z1} especially approaching
$z\sim2$ \citep{bouche:z2.kennicutt,shapiro:highz.kinematics,
  vanstarkenburg:z2.disk.dynamics}, and agrees well with the structure
of simulated disks undergoing rapid simultaneous accretion and
post-merger reformation
\citep{robertson.bullock:disk.merger.rem.vs.obs,
  governato:disk.rebuilding}.

Other observational signatures of this process include ``in situ''
bulge formation: rather than the conventional wisdom, in which most
classical (merger-induced) bulges are formed by mergers that destroyed
the progenitor disk at some early time, with a later re-accreted disk
around them, the dissipational models predict that it is much more
likely, especially in gas-rich (low-mass) systems, that such bulges
were formed by more recent mergers (many of them minor) that formed a
small bulge out of a large disk, without destroying most of the
disk. As such, the properties of the stellar populations in the bulge
should be reasonably continuous with those in the disk.  Note that, in
an integrated sense, the bulge may have older or more metal-rich
stars, but this is true even in a pure disk system if one compares the
central regions to the outer regions; moreover, star formation may
continue in the disk after the merger, whereas it may not in the
bulge. But, in terms of e.g.\ the full stellar population gradients
and even resolved color gradients, there should be continuity between
disk and bulge, with properties such as their metallicities
correlated. Indeed, high resolution resolved spectroscopic
observations of disk+bulge systems have begun to show that such
continuity is the norm, in metallicity and $\alpha$-enrichment,
stellar population ages, and colors 
\citep{moorthy:bulge.ssps, peletier:spiral.maps,
dominguez:bulge.vs.disk.colors,morelli:cluster.bulge.ssps,
macarthur:goods.bulge.ssps}.

Observations and simulations have shown that 
the two-component (gas+stellar) nature of mergers leads to a 
two-component structure of bulges, with a 
low-density outer component reflecting disk stars that are violently 
relaxed in the merger and dense inner component built by the 
gas that loses angular momentum \citep{mihos:cusps}. 
In a series of papers, \citet{hopkins:cusps.mergers,hopkins:cusps.ell,
hopkins:cores,hopkins:cusps.fp,hopkins:cusps.evol} 
demonstrated that with sufficiently high-dynamic range observations 
of spheroid structure and/or stellar populations, these 
can be decomposed, yielding empirical constraints on 
the efficiency of angular momentum loss in gas. 
Considering a large sample of ellipticals 
and recent merger remnants \citep{rj:profiles,lauer:bimodal.profiles,
jk:profiles}, the mass fraction of spheroids in the dissipational component 
appears, to lowest order, to trace typical disk gas fractions as a function of 
mass; however, \citet{hopkins:cusps.ell,hopkins:cusps.evol} note that 
at low masses, this appears to asymptote to a maximum 
dissipational/starburst fraction $\sim0.3-0.4$. Independent 
tests \citep{balcells:bulge.xl,
hopkins:cusps.mergers,jk:profiles} and constraints from stellar populations 
\citep{mcdermid:sauron.profiles,
sanchezblazquez:ssp.gradients,
reda:ssp.gradients} yield similar conclusions. 
If angular momentum loss in mergers were always efficient, 
the natural expectation would be that this should rise at low 
galaxy masses as the progenitor disk gas fractions, i.e.\ $\sim f_{\rm gas}\rightarrow1$. 
If, however, angular momentum loss becomes less efficient in these 
extremely gas-rich systems, the observed trend is the natural prediction: 
if the fraction of gas losing angular momentum scales as 
adopted here, then the dissipational fraction of the bulge formed 
from disks with gas fraction $f_{\rm gas}$ is not $\sim f_{\rm gas}$, 
but $\sim f_{\rm gas}/(1+f_{\rm gas})$, i.e.\ asymptoting 
to the values observed for all $f_{\rm gas}\sim0.5-0.9$. 
Expanded samples with high-resolution photometry, 
kinematics, and stellar population constraints would provide powerful 
constraints to extend these comparisons.

\acknowledgments 
We thank Fabio Governato, James Bullock, and Chris Purcell for helpful
discussions. We also thank the anonymous referee for 
suggestions that improved this manuscript. 
This work was supported in part by NSF grants ACI
96-19019, AST 00-71019, AST 02-06299, and AST 03-07690, and NASA ATP
grants NAG5-12140, NAG5-13292, and NAG5-13381.  Support for PFH was
provided by the Miller Institute for Basic Research in Science,
University of California Berkeley. Support for TJC was provided by the
W.~M.\ Keck Foundation. Support for DK was provided by Harvard
University and the Institute for Theory and Computation (ITC). RSS
thanks the ITC at Harvard University for supporting visits that
contributed to this work.
\\

\bibliography{/Users/phopkins/Documents/lars_galaxies/papers/ms}

\begin{appendix}

\section{Details of the Hopkins et al.\ 2009 (\disksurv) Model}
\label{sec:appendix}

The full gas fraction-dependent (``new'') model for the efficiency of
bulge formation as a function of gas fraction and merger properties,
discussed in the text, is derived in detail and compared to a large
suite of hydrodynamic merger simulations spanning the relevant
parameter space in \disksurv.  We refer the reader to that paper for a
detailed discussion of the physical motivation for the model and
numerical tests. Here, we briefly present the key physically-motivated
prescriptions which are shown therein to provide a good fit to the
simulation behavior and provide our implementation of the hydrodynamic
results in analytic or semi-analytic cosmological models.

Define: $m_{\ast,\,{\rm disk}}$ the mass of stars in the disk
component and $m_{\ast,\,{\rm bulge}}$ the mass of stars in the bulge
component, with $m_{\ast}=m_{\ast,\,{\rm disk}}+m_{\ast,\,{\rm
    bulge}}$ the total stellar mass; $m_{\rm cold}$ the mass of cold gas in
the galaxy (assumed to all reside in the disk). Here, $m_{\rm
  bar}=m_{\ast}+m_{\rm cold}$ is the baryonic mass of the galaxy, and
$r_{\rm d}$ is the effective radius of the disk. The disk gas fraction is
then $f_{\rm gas}=m_{\rm cold}/(m_{\rm cold}+m_{\ast,\,{\rm disk}})$
and the disk mass fraction is $f_{\rm disk} = (m_{\rm
  cold}+m_{\ast,\,{\rm disk}})/m_{\rm bar}$.\footnote{In simulations,
  the following descriptions are most robust if these quantities are
  defined just before the final coalescence/merger where the gas will
  lose angular momentum and stars will violently relax. Defining them
  at earlier times means that other quantities such as the star
  formation efficiency will be important for determining how much gas
  is actually present at the time of interest (the merger).}

The model outlined in \S~4.3 of \disksurv\ then gives a simple set of
formulae to compute the fraction of cold gas in both merger
progenitors that will participate in a central starburst (as a
consequence of losing its angular momentum), and as a result be 
removed from the disk and added to the bulge component of the
remnant. The mass of gas that participates in the starburst is
\begin{equation}
m_{\rm burst} \equiv m_{\rm cold}(< r_{\rm crit}), 
\label{eqn:mburst}
\end{equation}
or the mass of cold gas within a critical radius $r_{\rm crit}$ where
angular momentum loss from the gas disk in the merger is efficient.

\disksurv\ derive the following relation for $r_{\rm crit}$ (their
Equation~{(7)}):
\begin{equation}
\frac{r_{\rm crit}}{r_{\rm d}} = \alpha\,(1-f_{\rm gas})\,f_{\rm disk}\,F(\theta,b)\,G(\mu)
\label{eqn:rcrit}
\end{equation}
where 
\begin{equation}
F(\theta,b)\equiv {\Bigl (}\frac{1}{1+[b/r_{\rm d}]^{2}}{\Bigr)}^{3/2}\,
\frac{1}{1-\Omega_{\rm o}/\Omega_{\rm d}}
\label{eqn:Ftheta}
\end{equation}
contains the dependence on the merger orbit, and 
\begin{equation}
G(\mu)\equiv \frac{2\,\mu}{1+\mu}
\label{eqn:Gmu}
\end{equation}
contains the dependence on the merger mass ratio $\mu\equiv
m_{2}/m_{1}$. 
There is some ambiguity in the most relevant definition of 
``mass ratio'' \citep[e.g.][]{stewart:massratio.defn.conf.proc}; 
because we are interested in the gravitational dynamics induced 
by passage/merger, we find the best match to behavior in simulations 
by defining the masses of interest ($m_{1}$ and $m_{2}$) as 
the baryonic plus tightly bound (central) dark matter (specifically, 
dark matter mass within a core radius of $r_{s}$, the NFW scale length 
$\equiv r_{\rm vir}/c$, roughly $\sim10\%$ of $r_{\rm vir}$ or 
$\sim$a few disk scale lengths; the results are not strongly sensitive 
to the choice within a factor of $\sim2-3$ in $r$). 
This is important, in particular for low-mass galaxies which are observed 
to be dark-matter dominated (and have high $f_{\rm gas}$); simply 
considering the stellar or baryonic mass ratio significantly under-estimates 
the ``damage'' done to the disk relative to the results of high-resolution 
simulations. 

In Equation~\ref{eqn:Ftheta} above, $b$ is the distance
of pericentric passage on the relevant final passage before
coalescence ($\lesssim$ a couple $r_{\rm d}$ for typical cosmological
mergers), but \disksurv\ show that for systems which will actually
merge, as opposed to fly-by encounters, identical results are obtained
in the limit $b\rightarrow0$, since the final coalescence ---
regardless of initial impact parameter --- will be largely radial. The
quantity $\Omega_{\rm o}$ is the orbital frequency at pericentric
passage and $\Omega_{\rm d}$ is that of the disk;
\begin{eqnarray}
\nonumber \frac{\Omega_{\rm o}}{\Omega_{\rm d}} & = & \frac{v_{\rm peri}}{v_{c}}\,\frac{r_{\rm d}}{b}\,\cos{(\theta)}\\
& \approx & \sqrt{2\,(1+\mu)}\,[1+(b/r_{\rm d})^{2}]^{-3/4}\,\cos{(\theta)}, 
\end{eqnarray}
where $\theta$ is the inclination of the orbit relative to the disk 
($v_{\rm peri}$ is the velocity at pericentric passage, $v_{c}$ the 
circular velocity at the impact parameter $b$). 
We draw $\theta$ randomly for each merger assuming disks are
isotropically oriented (although this affects primarily the scatter,
not the mean relations predicted).

Adopting $b=2\,r_{\rm d}$, typical of cosmological mergers, or taking
the limit $b\rightarrow0$ for final radial infall gives a similar
scaling:
\begin{equation}
\alpha\,F(\theta,b) = \frac{0.5}{1-0.42\,\sqrt{1+\mu}\,\cos{(\theta)}}. 
\label{eqn:alphaF}
\end{equation}
The parameter $\alpha$ in Equation~(\ref{eqn:rcrit}) subsumes details of the 
stellar profile shape and bar/driven distortion dynamics in a merger. 
The precise value above is obtained from the numerical simulation results 
in \disksurv, but is also derived analytically therein. 

The disks are assumed to follow an exponential profile, so the
fraction of the cold gas mass within $r_{\rm crit}$ is given by
\begin{equation}
f_{\rm burst} = \frac{m_{\rm burst}}{m_{\rm cold}} 
= 1 - (1+ r_{\rm crit}/r_{\rm d})\,\exp{(-r_{\rm crit}/r_{\rm d})}. 
\label{eqn:fburst}
\end{equation}
Generally, evaluating these quantities with the cold gas and stellar
content being the sum of the two merging galaxies is a reasonable
approximation to the numerical results (representing the fact that the
burst takes place near or shortly after the time of coalescence); it
is also straightforward to independently apply the equations above to
both primary and secondary at the time of the merger. In practice, it
makes little difference (even assuming that the entire gas mass of the
secondary is part of the burst is a good approximation for all but the
most rare very major mergers with mass ratios close to 1:1).

Following \disksurv\ \S~4.3.3, we further assume that the mass of the
stellar disk in the primary (more massive) galaxy that is transferred
to the bulge/spheroidal component from the disk and violently relaxed
is
\begin{equation}
m_{\rm disk,\,destroyed} = \mu\,m_{\ast,\,{\rm disk}}.
\label{eqn:mdiskdestroyed}
\end{equation}
Furthermore, the whole stellar mass of the secondary, $m_{2}$, is also added
to the spheroidal remnant.  A more detailed accounting of e.g.\ the
efficiency of violent relaxation as a function of radius \citep[see
  \disksurv,][]{hopkins:disk.heating} in the stellar disk suggests
that one should further multiply the right-hand side of
Equation~\ref{eqn:mdiskdestroyed} by a factor $(1+\mu^{-0.3})^{-1}$;
in practice including this weak dependence on $\mu$ makes no
difference.

As described in the text, the gas
fraction-independent or ``dissipationless'' model treats gas in the same way as stars in
terms of the efficiency of merger-induced bulge formation. Gas which
is assumed to be transferred to the spheroid in the merger is still
allowed to lose angular momentum and participate in a central
starburst as opposed to being violently relaxed (so the system is not,
formally speaking, dissipationless), but the gas mass fraction that
does so scales with merger properties in the same manner as the
dissipationless violently relaxed stellar fraction.  This 
amounts to replacing our more physical model for the gas behavior with
the analogue of Equation~\ref{eqn:mdiskdestroyed}: the gas mass which
goes into a starburst in the merger is simply $m_{\rm burst} =
\mu\,m_{\rm cold}$, independent of gas fraction. Since the behavior of
interest is the dependence of bulge formation efficiency on gas
fraction, we can instead still choose to adopt Equation~\ref{eqn:rcrit}
for the behavior of the gas, but omit the dependence on disk
gas fraction (the $(1-f_{\rm gas})$ term), equivalent to assuming that
the gas always loses angular momentum as efficiently as in a
relatively gas-poor merger (formally computing
Equation~\ref{eqn:rcrit} with $f_{\rm gas}\rightarrow0$, regardless of
the actual disk gas fraction).  This choice makes no difference in our
calculations and comparisons throughout the paper: the dependence on
e.g.\ orbital parameters in Equation~\ref{eqn:rcrit} only affects the
scatter in $B/T$, not the mean trends, and the scaling with mass ratio
is very similar.  The important difference between the models is
entirely contained in the $(1-f_{\rm gas})$ scaling.

We emphasize that there are no free or ``tuned'' parameters in this
model. All of the parameters and values given above are determined in
\disksurv\ based on physical models and hydrodynamic merger
simulations, and are adopted without adjustment in both the empirical
and the semi-analytic model.

\section{Details of the S08 Semi-analytic Model} 
\label{sec:appendix:S08}

Here, we give further details of the semi-analytic model in which 
we embed the above models for bulge formation. This model has been
described in detail in \citet{somerville99:sam,somerville:sam} and a
version with major updates has recently been described in
\citet[][S08]{somerville:new.sam}. We refer the reader to
those papers for details, but briefly summarize the main ingredients
of the models here.  The models are based on Monte Carlo realizations
of dark matter halo merger histories, constructed using a slightly
modified version of the method of \citet{somerville:merger.trees}, as
described in S08. The merging of sub-halos within virialized dark
matter halos is modeled by computing the time required for the
subhalo to lose its angular momentum via dynamical friction and fall
to the center of the parent halo. The S08 model takes into account the
mass loss of the orbiting dark matter halos owing to tidal stripping, as
well as tidal destruction of satellite galaxies.  

The model discriminates between cooling and accretion in two different
regimes: ``cold mode'', in which the gas cooling time is less than the
free-fall time, and ``hot mode'', in which the opposite condition
holds. In the cold mode, gas accretion onto the disk is limited only
by the rate at which gas is incorporated into the virialized halo. In
the hot mode, the rate at which gas can lose its energy and cool via
atomic cooling is estimated.  This cooling gas is assumed to initially
settle into a thin exponential disk, supported by its angular
momentum. Given the halo's concentration parameter, spin parameter,
and the fraction of baryons in the disc, angular momentum conservation
arguments are used to compute the scale radius of the exponential disc
after collapse \citep{somerville:disk.size.evol}.  The models make use
of fairly standard recipes for the suppression of gas infall owing to a
photo-ionizing background, star formation, and supernova feedback, as
described in S08. Quiescent star formation, in isolated discs, is
modeled using the empirical Kennicutt star-formation law
\citep{kennicutt98}. A critical surface density threshold is adopted,
and only gas lying at surface densities above this value is available
for star formation. Cold gas may be reheated and ejected from the
galaxy, and possibly from the halo, by supernova feedback. Heavy
elements are produced with each generation of stars, and enrich both
the cold and hot gas.  

Galaxy mergers trigger the accretion of gas onto supermassive black
holes in galactic nuclei. Energy radiated by black holes during this
`bright', quasar-like mode can drive galactic-scale winds, clearing
cold gas from the post-merger remnants. In addition to the rapid
growth of BHs in the merger-fueled, radiatively efficient `bright
mode', BHs also experience a low-Eddington-ratio, radiatively
inefficient mode of growth associated with efficient production of
radio jets that can heat gas in a quasi-hydrostatic hot halo. Thus,
once the black hole has grown sufficiently massive, this ``radio
mode'' feedback can completely switch off further cooling, eventually
quenching star formation in massive galaxies. Note that the radio mode
feedback is assumed only to be effective in heating gas that is being
accreted in the ``hot mode'' described above.  The semi-analytic
models contain a number of free parameters, which are summarized in
Table~1 of S08. These parameters were chosen in order to reproduce
certain observations of nearby galaxies, as described in detail in S08
(most importantly, the stellar mass function, cold gas fraction of
spirals as a function of stellar mass, and the metallicity-mass
relation zero-point). In this work we use the fiducial values of all
parameters, as given in S08.

\section{Variations of the Semi-Empirical Model} 
\label{sec:appendix:variations}

An advantage of the semi-empirical model approach is that it is 
trivial to alter various assumptions, provided that
the key observational constraints (e.g.\ galaxy masses and gas fractions, in 
this case) remain satisfied. We have experimented with a number of 
such variations, and find that the conclusions in this paper are robust. 
A detailed discussion of these variations and their consequences for 
merger rates and galaxy merger histories will be presented in 
future work \citep{hopkins:merger.rates}. Here, we briefly outline some of these 
choices. 

Changing the adopted dark matter merger rates makes little difference. 
We have re-calculated our model adopting the halo merger 
trees from \citet{stewart:merger.rates}, using the subhalo-based 
merger trees from \citet{wetzel:mgr.rate.subhalos}, 
and implementing various approximations for a ``merger timescale'' 
(delay between halo-halo and galaxy-galaxy merger, given 
e.g.\ by some approximation to the dynamical friction time) 
from \citet{boylankolchin:merger.time.calibration} and 
\citet{jiang:dynfric.calibration}. We have also re-calculated the model 
using different observational constraints to determine the 
galaxy masses populated in halos or subhalos of a given 
mass; using the monotonic-ranking method of \citet{conroy:monotonic.hod} 
applied to the redshift-dependent galaxy mass functions 
from \citet{fontana:highz.mfs} or \citet{perezgonzalez:mf.compilation}, 
or using the $z=0$ constraints determined directly from clustering of 
SDSS galaxies in \citet{wang:sdss.hod} and ignoring redshift evolution 
in this quantity. We find in each case a similar result. 

We can also change the methodology, to essentially a toy model 
(between our strict semi-empirical approach and a semi-analytic model) 
fitted to observations. We assume at each timestep that 
all galaxies in halos below $10^{12}\,\msun$ simply 
accrete cold gas proportional to their halo growth 
\citep[this corresponding to what is seen in simulations and 
observations; low mass galaxies accrete efficiently in the ``cold mode'' 
with gas coming directly into disks on a dynamical time, high 
mass galaxies are ``quenched''; see e.g.][]{keres:hot.halos,keres:cooling.revised}.
We can then determine a star formation rate either by 
forcing the galaxies to move along the empirically constrained 
$M_{\rm gal}(M_{\rm halo})$ track (our default method, based on the observational 
constraints described above), or by assuming the 
galaxies lie on a fixed size-mass relation and using the \citet{kennicutt98} 
relation \citep[both appear to evolve only weakly with redshift; 
see][]{bouche:z2.kennicutt,puech:tf.evol,
genzel:highz.rapid.secular,vanstarkenburg:z2.disk.dynamics}, or 
by assuming they live on the fitted SFR-galaxy mass relation in \citet{noeske:sfh}. 
We can further 
allow some ``outflow,'' with an efficiency $\dot{M}_{\rm outflow} \propto 
\dot{M}_{\ast}$ fitted so as to yield a good match to 
the $f_{\rm gas}$-galaxy mass relation observed 
at different redshifts \citep[e.g.][]{belldejong:disk.sfh,mcgaugh:tf,
shapley:z1.abundances,erb:lbg.gasmasses,calura:sdss.gas.fracs}. 
Recall, in our ``default'' semi-empirical 
model, we simply assume whatever inflow/outflow 
necessary such that the galaxies always live on the $f_{\rm gas}$-galaxy 
mass relation fitted to these observations; but appropriately tuned, this 
simple toy model can yield a similar result. 

Ultimately, we find that these choices make little difference, so long 
as they yield the same distribution of galaxy masses and galaxy 
gas fractions. Those are the parameters that effect the model conclusions; 
to the extent that the variations above provide a means to 
interpolate between the empirical constraints on these quantities, they 
yield similar conclusions. 
  
\end{appendix}

\end{document}